\documentclass[titlepage,11pt]{article}

\usepackage{graphicx}

\usepackage[myheadings]{fullpage}
\usepackage{pmetrika}
\usepackage{apacite}
\usepackage{submit}
\usepackage{amsmath,amsfonts,amssymb}
\usepackage{algorithm}
\usepackage{algpseudocode}
\usepackage{array,ragged2e}
\usepackage{booktabs,multirow}
\newcolumntype{P}[1]{>{\RaggedRight\arraybackslash}p{#1}}
\newcolumntype{C}[1]{>{\centering\arraybackslash}p{#1}}

\def\vec{\mathrm{vec}}

\def\xx{\mathbf{x}}

\def\vv{{\mathbf V}}

\def\yy{\mathbf{y}}
\def\YY{\mathbf{Y}}
\def\WW{\mathbf{W}}
\def\ss{\mathbf{s}}

\def\ssum{\sum_{j=1}^J}
\def\yyi{\mathbf{y}_i}
\def\XX{\mathbf{X}}
\def\ggamma{\bfgamma}
\def\II{\boldsymbol{\mathcal{I}}}
\def\SSigma{\bfSigma}
\def\ddelta{\bfdelta}
\def\eeta{\bfeta}
\def\bbeta{\bfbeta}
\def\xxi{\bfxi}
\def\ppsi{\bfpsi}
\def\real{\mathcal{R}}
\DeclareMathOperator*{\argmin}{\arg\min}
\DeclareMathOperator{\E}{\mathbb{E}}

\DeclareRobustCommand{\bigO}{%
  \text{\usefont{OMS}{cmsy}{m}{n}O}%
}

\begin{document}

\begin{titlepage}

\title{STATISTICAL INFERENCE FOR THE PENALIZED EM ALGORITHM TO TEST DIFFERENTIAL ITEM FUNCTIONING}


\author{Weimeng Wang, Yang Liu, and Jeffrey R. Harring}

\affil{University of Maryland, College Park}


\vspace{\fill}\centerline{\today}\vspace{\fill}

\linespacing{1}
\contact{Correspondence should be sent to Weimeng Wang\\
\noindent E-Mail: weimengbonnie@gmail.com \\
\noindent Affiliation: University of Maryland, College Park
}
\end{titlepage}

\setcounter{page}{2}
\vspace*{2\baselineskip}


\linespacing{1.5}
\abstracthead
\begin{abstract}
\noindent
Recent advancements in testing differential item functioning (DIF) have greatly relaxed restrictions made by the conventional multiple group item response theory (IRT) model with respect to the number of grouping variables and the assumption of predefined DIF-free anchor items. The application of the $L_1$ penalty in DIF detection has shown promising results in identifying a DIF item without a priori knowledge on anchor items while allowing the simultaneous investigation of multiple grouping variables. The least absolute shrinkage and selection operator (LASSO) is added directly to the loss function to encourage variable sparsity such that DIF parameters of anchor items are penalized to be zero. Therefore, no predefined anchor items are needed. However, DIF detection using LASSO requires a non-trivial model selection consistency assumption and is difficult to draw statistical inference. Given the importance of identifying DIF items in test development, this study aims to apply the decorrelated score test \cite{ning2017general} to test DIF once the penalized method is used. Unlike the existing regularized DIF method which is unable to test the statistical significance of a DIF item selected by LASSO, the decorrelated score test requires weaker assumptions and is able to provide asymptotically valid inference to test DIF. Additionally, the deccorrelated score function can be used to construct asymptotically unbiased normal and efficient DIF parameter estimates via a one-step correction. The performance of the proposed decorrelated score test and the one-step estimator are evaluated via a Monte Carlo simulation study.
\end{abstract}
\vspace{\fill}\pagebreak

\section{Introduction}

A major goal of educational and psychological assessment is to create interpretable scores on a relevant construct being measured. A well-constructed test score representing a latent trait (e.g., proficiency level), should be reliable and valid. One major threat to the validity of the intended use of scores from an instrument is the lack of measurement invariance or existence of {\it differential item functioning} (DIF). Sometimes, characteristics of a test may introduce unintended systematic score differences between individual test takers or subgroups of examinees of the same ability level, which can result in invalid score interpretation. Therefore, DIF detection should be conducted for routine operations of psychological and educational assessment to ensure valid interpretations of test scores.

Item response theory (IRT), a collection of mathematical and statistical models, was\textemdash and still remains\textemdash the foundational modeling system in educational assessment \cite{Embretson}. One advantage of an IRT model in conducting a DIF analysis is that it offers a formal statistical framework for assessing DIF. Despite the popularity and advantages of DIF detection using IRT models, the inherent assumptions therein make practical applications of DIF analyses challenging. For example, IRT DIF detection methods usually explicitly or implicitly assume the existence of {\it predefined} DIF-free items\footnote{See \citeA{wangliuliu2022} for an exception.}, also known as anchor items. Typically, DIF-free items are needed to anchor the latent scale so that DIF can be distinguished from the between group difference in latent trait distributions. Correctly specifying anchor items is crucial in correctly identifying a DIF item. Violations to the DIF-free anchor assumption may lead to inflated false detection rates in finding DIF items \cite{wang2004effects,wang2003effects,woods2009empiricaldif,woodscaiWang}. A practical obstacle is locating DIF-free anchor items without a priori information. Another challenge is detecting DIF items associated with multiple grouping variables. Conventional IRT DIF detection methods typically investigate DIF with respect to one categorical grouping variable (e.g., gender, ethnicity, SES) at a time. However, it is impossible to investigate the complex nature of DIF due to interconnected background characteristics \cite{liu2017differential} and thus it is unlikely to truly uncover the source of DIF \cite{shea2013using}. 

As a solution to these practical difficulties encountered by conventional IRT DIF methods, regularization has been applied to the latent regression modeling framework to detect DIF items. In particular, the use of the $L_1$ penalty or the LASSO has shown promising results in DIF detection due to its efficiency in variable selection \cite<e.g., >{bauer2017more,belzak2020improving,magis2015detection,tutz2015penalty}. For example, the $L_1$ penalty has been successfully applied to moderated nonlinear factor analysis (MNLFA) model to identify DIF items \cite{bauer2020simplifying,belzak2020improving}. Such a modeling framework offers greater flexibility to detect DIF for multiple grouping variables simultaneously. Item parameters and population distribution parameters, upon applying suitable link functions, can be expressed as linear or nonlinear functions of multiple person characteristics. Consequently, DIF detection can be treated as a variable selection problem---non-zero coefficients are only assigned to covariates that cause DIF. The goal of the penalty procedure is to encourage sparseness as much as possible with respect to DIF parameters such that DIF parameters of anchor items are penalized to be zero while those of DIF items are non-zero. Hence, no predefined anchor items are needed as items with DIF parameters penalized to be zero are used as anchors. 

Despite the flexibility of the $L_1$ penalty for DIF detection, several concerns need to be addressed carefully. First, the accuracy of DIF detection based on the $L_1$ penalty is contingent on non-trivial assumptions. In order for LASSO to select the right covariate for each item, and thus, items exhibiting true DIF, a variable selection consistency condition is needed \cite{zhao2006model}. In the current case, failing to meet this condition could result in mistakingly flagging non-DIF items or miss real DIF items even in large samples. Second, with a finite sample size especially those encountered in the social sciences, inferential statistics such as confidence interval estimates and p-values are extremely important to differentiate a true DIF item from a sampling error. For example, a common practice in DIF detection in educational and psychological assessments is effect size reporting after DIF is detected. In this case, an uncertainty measure may prove to be helpful in deciding whether a DIF effect size is truly different from zero. Nevertheless, it is difficult to draw statistical inference based on the LASSO-type estimator due to shrinkage resulting from penalization \cite{belzak2020improving,huang2018penalized,lindstrom2020model,tutz2015penalty}. Lastly, LASSO creates biased parameter estimates and non-normal limiting distribution of the parameter estimates\cite{fu2000asymptotics}, which makes subsequent analyses difficult \cite{chen2021advantages,fan2001variable}. For example, DIF effect sizes, item parameter estimates, and standard error estimates cannot be directly used to quantify uncertainty in latent score estimates \cite{liu2018bootstrap}. 

\section{Purpose of the Study}

The current study aims to fill the gap in the literature to make inferential claims once LASSO is used. Specifically, a decorrelated score test \cite{ning2017general} is proposed to detect DIF for binary response data with multiple covariates after the $L_1$ penalty is used. The decorrelated score test could potentially be extended to accommodate continuous and discrete response data and multidimensional latent variables. In the current study, the simplest and the most commonly used case was considered to better demonstrate the extension of the decorrelated score test to detect DIF. Unlike the existing regularized DIF method, the decorrelated score test does not require the variable selection consistency assumption and is able to provide valid inference on DIF effects. Specifically, a sparse score vector with respect to the focal parameter is estimated consistently so that the resulting score test statistic has an asymptotically normal reference distribution. Additionally, an asymptotic unbiased estimator can be constructed using a one-step bias correction using the decorrelated score function.

A Monte Carlo simulation study is conducted to examinee the finite sample behavior of the decorrelated scodre test. Its performance in controlling the Type I error rate, establishing sufficient power, controlling the false detection rate in identifying a DIF item under different DIF-related conditions are compared with three methods: (1) regularization method based on LASSO selection only, (2) a naive model refitting method \cite{belzak2020improving}, and (3) the oracle solution assuming known anchors. 

\section{Decorrelated Score Test to Test DIF}\label{s:difdefinition}
Mathematically, DIF can be defined as follows. Considering $n$ independent and identically distributed multivariate vectors, $\mathfrak{Y}=(\YY_1,\cdots,\YY_n)^\top$ and $\YY_i=(Y_{i1},\dots, Y_{iJ})^\top$, following a statistical model $\mathcal{P}=\{P_{\xxi}:\xxi \in \Xi\}$, where $\xxi$ is a $d-$dimensional vector of unknown parameters and $\Xi \subset \real^d$ is the parameter space. The categorical random variable $Y_{ij}$ representing the item response from person $i$, $i=1,\cdots, n$ to item $j \in \mathcal{J}=\{1,\dots,J\}$, define the conditional item response function, denoted as $f_j(y|\theta,\xx)=P(Y_{ij}=y|\theta_i=\theta,\xx_i=\xx)$, as the probability of endorsing a particular answer conditional on the person's latent ability level $\theta_i=\theta\in\real$ and person covariate $\xx_i\in\real^K$. More
broadly, item $j$ exhibits DIF if
\begin{align} \label{eq: nonpar-dif}
	f_j(y|\theta,\xx) \neq f_j(y|\theta, \tilde{\xx}),
\end{align}
for some $\xx \neq \tilde{\xx}\in\real^K$. Equation
\ref{eq: nonpar-dif} indicates that DIF exists when the conditional item
response function $f_j$ differs for any $\xx \neq \tilde{\xx}$ after
controlling for the same level of latent ability. In other words,
measurement invariance holds for item $j$, only when $ f_j(y|\theta,\xx) =
f_j(y|\theta, \tilde{\xx})$ for all $\xx \neq \tilde{\xx}$. 
\subsection{Moderated Nonlinear Factor Analysis Model}

The moderated nonlinear factor analysis model (MNLFA) assumes that the item response for item $j$ from person
$i$, (i.e., $Y_{ij} \in \{0, 1\}$) follows a Bernoulli distribution ($ Y_{ij}|\theta_i
\sim \hbox{Bern}(P_{ij}(\theta_i))$), where $P_{ij}$ denotes the probability of endorsing
item $j$ for person $i$. Incorporating vector of person covariates $\xx_i$ into the
conditional item response function (cIRF) to affect both the item response and
the latent distribution offers greater flexibility in testing DIF across different
levels of both continuous and categorical variables. It also permits testing
DIF for multiple grouping variables as well as their
interaction effects (e.g., gender $\times$ age) at the same time. Using the same notation, the cIRF
$f_j$ is written as
\begin{align}\label{eq:cIRF}
 	f_j(Y_{ij}|\theta,\xx)&=P_{ij}^{Y_{ij}}(1-P_{ij})^{1-Y_{ij}},\,
\hbox{where}\nonumber\\
 	P_{ij}=P(Y_{ij}&=1|\theta_i=\theta,
\xx_i=\xx)=\frac{1}{1+\exp[-\alpha_j(\xx)-\beta_j(\xx)\theta]},
 \end{align}
 where $\alpha_j(\xx)$ and $\beta_j(\xx)$ are the item intercept and item slope functions,
respectively, which can be further expressed as functions of the vector of person covariate
$\xx\in \real^{K}$. Latent ability $\theta_i$ is assumed to be
normally distributed: $\theta_i\sim \mathcal{N}(\mu(\xx),\sigma^2(\xx))$, in
which the latent population parameters: mean ($\mu$) and variance ($\sigma^2$)
can also be expressed as functions of the person covariate vector $\xx$ defined as follows:
 \begin{align} \label{eq:blatent}
	\mu(\xx)=\ggamma^\top\xx \quad \sigma^2(\xx)=\exp(\ddelta^\top\xx),
\end{align}
where $K$-dimensional vectors $\ggamma$ and $\ddelta$ represent effects of the
person covariate vector $\xx$ on the latent mean and variance on a logarithmic scale. These two vectors are further referred to as latent population parameters. Also, item intercept and item slope functions are expressed as
 \begin{align}
 	\alpha_j(\xx)&=d_j+\bbeta_{0j}^\top\xx, \label{eq:unif} \\
 	\beta_j(\xx)&=a_j+\bbeta_{1j}^\top\xx \label{eq:nonunif},
 \end{align}
where $d_j$ and $a_j$ are the item location and item slope parameters. The $K$-dimensional vectors $\bbeta_{0j}$ and $\bbeta_{1j}$ stand for person
covariate effects on the item intercept and item slope parameter, respectively.
To detect DIF for item $j$, a composite hypothesis test is typically conducted to test against the null hypothesis
 \begin{align} \label{eq:null}
 	\hbox{H}_0: \,  \bbeta_{0j}=\bbeta_{1j}=\bf{0}.
 \end{align}
Typically, a minimal constraint is needed to identify the MNLFA model. This requires that there exists at least one item for each column of $\bbeta_{0j}$ and $\bbeta_{1j}$ is zero/anchor. Without predefined anchor items or constrained parameters, the MNLFA model can be estimated using the Penalized Expectation Maximization (EM) Algorithm and DIF items are those items with remaining DIF parameters.

\subsection{Regularized Differential Item Functioning}
The marginal likelihood
function $f(\yyi|\xx_i)$ of an observed individual response vector $\yyi=(y_{i1},\dots,y_{1J})^\top$
assuming conditional independence among item responses given the latent variable and person covariates $\xx_i=(x_{i1},\dots,x_{iK})^\top$ is
expressed as
 \begin{align} \label{eq:marginal}
 	f(\yyi|\xx_i)=\int\prod_{j=1}^J f_j(y_{ij}|\theta_i,\xx_i)\phi(\theta_i|\xx_i)d\theta_i,
 \end{align}
 in which $\phi(\cdot)$ is the probably density function of a normal
distribution governed by the population parameters $\ggamma$ and $\ddelta$. Assuming that the item response vectors
$\mathfrak{Y}$ is independent and identically
distributed (i.i.d), the sample likelihood function is written as
 \begin{align} \label{eq:likelihood}
 	f_n(\YY|\XX)=\prod_{i=1}^nf(\yyi|\xx_i),
 \end{align}
 where $\YY=(\yy_1,\cdots,\yy_n)^\top$ and $\XX=(\xx_1,\cdots,\xx_n)^\top$ represent observed data matrices.
 Subsequently, the corresponding marginal log-likelihood is expressed as
  \begin{align} \label{eq:ln}
 \ell_n(\xxi; \YY|\XX)=\frac{1}{n}\log f_n(\YY|\XX)=\frac{1}{n}\sum_{i=1}^n\log f(\yyi|\xx_i),
 \end{align}
 where $\xxi = (a_1, d_1, \bbeta_{01}^\top, \bbeta_{11}^\top, \dots, a_J, d_J, \bbeta_{0J}^\top, \bbeta_{1J}^\top,\ggamma^\top,\ddelta^\top)^\top$ is a collection of model parameters. To estimate the model using the penalized EM algorithm, the $L_1$ penalty function is added directly to the loss function 
\begin{align}  \label{eq:pmlikn}
  p_n(\xxi; \YY, \lambda|\XX) =-\ell_n(\xxi;\YY|\XX)+\lambda\|\bbeta_j\|_1,
  \end{align}
  where $\bbeta_j=(\bbeta_{0j}^\top,\bbeta_{1j}^\top)^\top$ denotes all DIF parameters, $\ell_n(\xxi;\YY|\XX)$ is the sample log-likelihood expressed in Equation \ref{eq:ln} and $\|\|_1$ indicates the $L_1$ norm that sums absolute values of all elements weighted by the penalty weight $\lambda$. Then the local minimizer of $ p_n(\xxi; \YY, \lambda|\XX)$ with respect to $\xxi$ 
 \begin{align}
  \hat{\xxi} &= \argmin_{\xxi} p_n(\xxi;\YY,\lambda|\XX)
\end{align}
can be estimated by the penalized EM algorithm, which is documented in Appendix A. 
 
In the upcoming section, the decorrelated score test will be introduced. The intuition about the decorrelated score test resembles the efficient score test when the model is identified. Thus, the efficient score test for an identified model will first be introduced followed by the decorrelated score test when the model is not identified with redundant parameters. 
\subsection{Decorrelated Score Test} \label{ss:dscoretest}
Partition model parameters $\xxi$ into $\xxi =(\ppsi,\eeta^\top )^\top$, where $\ppsi$ is the $d_0-$dimensional vector of focal parameters and $\eeta^\top $ stands for $d_1=(d-d_0)-$dimensional vector of nuisance parameters. Given the loss function $\ell(\xxi,\YY)=-\ell_n(\xxi,\YY)$, which is the average sample log-likelihood, define $\II=\E_{\xxi}(\nabla^2\ell(\xxi,\YY))$. Let $\xxi^*$ be the true value of $\xxi$. Similarly, denote $\II^*=\E_{\xxi^*}(\nabla^2\ell(\xxi^*,\YY))$. Note that for the rest of the study, $d$ is fixed. Let's first consider the case when the model is identified. For example, an MNLFA model with sufficiently large subset of known anchors. The asymptotic normality of the ML estimator [$\hat{\xxi}=\hbox{argmin}_{\xxi \in \Xi}\ell(\xxi,\YY)$] follows from a first-order Taylor series expansion of the score function at $\xxi^*$, 
\vspace{-12pt}
\begin{align*}
	\bold{0}&=\nabla_{\xxi}\ell(\hat{\xxi},\YY)=\nabla_{\xxi}\ell({\xxi^*},\YY)+\nabla^2_{\xxi,\xxi}\ell({\xxi^*},\YY)(\hat{\xxi}-\xxi^*)+R_n
\end{align*}
where the remainder term is of the form $R_n=\frac{1}{2}(\hat{\xxi}-\bar{\xxi})^T\nabla_{\xxi,\xxi}^3\ell(\bar{\xxi},\YY)(\hat{\xxi}-\bar{\xxi})$ and $\bar{\xxi}$ is in between $\hat{\xxi}$ and $\xxi^*$. By rearranging this equation,
\begin{align*}
 \sqrt{n}\,(\hat{\xxi}-\xxi^*)&=-\sqrt{n}\,\II^{*-1}\nabla_{\xxi}\ell({\xxi^*},\YY)+\sqrt{n}R_n\\
	\Leftrightarrow \sqrt{n}\,\begin{pmatrix}
		\hat{\ppsi} -\ppsi^*\\ \hat{\eeta} -\eeta^*
	\end{pmatrix}&=-\sqrt{n}\,\begin{pmatrix}
	\II^*_{\ppsi,\ppsi}& \II^*_{\ppsi,\eeta}\\
	\II^*_{\eeta,\ppsi}&\II^*_{\eeta,\eeta}
	\end{pmatrix}^{-1}\begin{pmatrix}
	\nabla_{\ppsi}\ell({\xxi^*},\YY)\\
	\nabla_{\eeta}\ell({\xxi^*},\YY)
	\end{pmatrix}-\sqrt{n}R_n,
\end{align*}
where subscripts of $\II^*$ are corresponding partitions of the matrix. Drop $\YY$ from the rest of the notations. As $\II_{\eeta,\eeta}$ is invertible, using the block inverse formula, it follows that
\begin{align*}
	 \sqrt{n}\,(\hat{\ppsi}-\ppsi^*)=-\sqrt{n}\,\II^{*-1}_{\ppsi|\eeta}s({\xxi^*})+\sqrt{n}R_n,\quad R_n= o_p(1/\sqrt{n})
	\end{align*}
where $\ss({\xxi^*})=\nabla_{\ppsi}\ell(\xxi^*)-\II^*_{\ppsi\eeta}\II^{*-1}_{\eeta\eeta}\nabla_{\eeta}\ell(\xxi^*)$ and $\II^*_{\ppsi|\eeta}=\II^*_{\ppsi\ppsi}-\II^*_{\ppsi\eeta}\II^{*-1}_{\eeta\eeta}\II^*_{\eeta\ppsi}$ are the efficient score and efficient information, respectively. Note that efficient score can be interpreted as the projection of the $\nabla_{\ppsi}\ell(\xxi)$ to the orthogonal complement of the score function with respect to the nuisance parameters \cite{vaart_1998}. Under $H_{0}: \ppsi^*=\mathbf{0}$, it holds that $n\nabla_{\ppsi}(\mathbf{0},\hat{\eeta})^{\top}[\hat{\II}_{\ppsi|\eeta}]^{-1}\nabla_{\ppsi}(\mathbf{0},\hat{\eeta})\overset{\mathcal{D}}{\rightarrow}\chi^2_{d_0}$, where $\hat{\eeta}=\hbox{argmin}_{\eeta}\ell(\mathbf{0},\eeta)$ is constrained parameter estimates. It is straightforward to see that the estimated efficient score ($\ss(\mathbf{0},\hat{\eeta})$) equals $\nabla_{\ppsi
}(\mathbf{0},\hat{\eeta})
$ due to the fact that $\nabla_{\eeta}\ell(\mathbf{0},\hat{\eeta})=\mathbf{0}$.

However, when the model is not identified, $\II_{\eeta,\eeta}$, in general, is not invertible. Intuitively, there exists redundant parameters in the model. A natural extension of the above case when the model is not identified is estimating an ``efficient score'' so that the influence of entries of the redundant parameters in the nuisance score is minimized. Following the same logic as the efficient score, a sparse score can be estimated by projecting $\nabla_{\ppsi}\ell(\xxi)$ to the orthogonal complement of a low-dimensional subspace spanned by the non-redundant part of the nuisance score vector. Therefore, a sparse vector/matrix is needed to find the best sparse linear combination of $\nabla_{\eeta}\ell(\xxi)$ to approximate $\nabla_{\ppsi}\ell(\xxi)$. Mathematically, the decorrelated score, or the extension of the efficient score when the model is not identified, has no correlation with the score function with respect to the nuisance score (i.e., $\E(\ss(\xxi)^\top\nabla_{\eeta}\ell(\xxi))=\mathbf{0}$). Geometrically, the decorrelated score has the same interpretation which is the projection of $\nabla_{\ppsi}\ell(\xxi)$ to the orthogonal complement of the linear space spanned by the nuisance score function $\nabla_{\eeta}\ell(\xxi)$. Following this intuition, define the decorrelated score function as
\begin{align} \label{eq:dscore}
\ss(\xxi) & = \nabla_{\ppsi}\ell(\xxi)-\WW^\top\nabla_{\eeta}\ell(\xxi), \hbox{ where } \WW^{\top}= \II_{\ppsi\eeta}\II^{-1}_{\eeta\eeta} \in \real^{d_0\times d_1}.
 \end{align}
To find the projection of $\nabla_{\ppsi}\ell(\xxi)$ to the orthogonal complement of the linear space span by the nuisance score function $\nabla_{\eeta}\ell(\xxi)$, the sparse matrix $\WW$ can be estimated using Algorithm \ref{alg:dscore} (see Table \ref{tab:algorithm1}). As can be seen, the key is to estimate a sparse matrix $\hat{\WW}=(\hat{\WW}_{*1},\cdots,\hat{\WW}_{*d_0})$ column by column to construct the decorrelated score ($\ss(\ppsi,\eeta)$) so that the additional and redundant parameters do not influence the decorrelated score.
\tablehere{1}
\noindent Further the Dscore test statistic can be constructed
\begin{align}
	\hat{T}_{Dscore}=n\hat{\ss}(\mathbf{0},\eeta)^{\top}[\hat{\II}_{\ppsi|\eeta}]^{-1}\hat{\ss}(\mathbf{0},\eeta),
\end{align}
where $\hat{\II}_{\ppsi|\eeta}=\nabla^2_{\ppsi,\ppsi}\ell(\hat{\xxi})-\hat{\WW}^{\top}\nabla^2_{\eeta,\ppsi}\ell(\hat{\xxi})$.
Under some technical assumptions\footnote{These assumptions are documented in the Appendix \ref{ch:Appendixb} and are not yet verified in the latent variable models. The intuition of the assumption verification is discussed in the Appendix \ref{ch:Appendixb}}, it can be proved that 
\begin{align}\label{eq:Fdscore}
	\hat{T}_{Dscore}\xrightarrow{\cal{D}} \chi^2_{d_0}.
\end{align}
Moreover, a one-step asymptotical unbiased estimator $\tilde{\ppsi}$ can be constructed using a single Newton step as 
\begin{align} \label{eq:onestepubiased}
	\tilde{\ppsi}=\hat{\ppsi}-\hat{\II}^{-1}_{\ppsi|\eeta}\hat{\ss}(\hat{\xxi})
\end{align}
and with assumptions \ref{asp:1} to \ref{asp:4} documented in Appendix \ref{ch:Appendixb}, \citeA{ning2017general} showed that
\begin{align} \label{eq:Funbiased}
\sqrt{n}(\tilde{\ppsi}-\ppsi^*)\xrightarrow{\cal{D}}{{\cal N}(0,\II_{\ppsi|\eeta}^{*-1})}.
\end{align}
Subsequently, the $(1-\alpha)\times100\%$ confidence interval can be constructed for a linear combination of $\ppsi^*$ (i.e., $\mathbf{c}^{\top}\ppsi^*$ where $\mathbf{c}^{\top}$ is a $d_0-$dimensional constant vector) as $[\tilde{\ppsi}-n^{-1/2}\Phi^{-1}(1-\alpha/2)(\mathbf{c}^\top\hat{\II}_{\ppsi|\eeta}\mathbf{c})^{-1/2},\tilde{\ppsi}+n^{-1/2}\Phi^{-1}(1-\alpha/2)(\mathbf{c}^\top\hat{\II}_{\ppsi|\eeta}\mathbf{c})^{-1/2}]$.

\section{Monte Carlo Simulation} \label{s:simulation}
In order to investigate the finite sample behavior of the proposed Dscore test in testing DIF under different conditions, a Monte Carlo study is conducted. Performance of the proposed Dscore test is evaluated in comparison to three methods: (1) the Wald test assuming known and correctly specified anchors, (2) the Reg-DIF method (i.e., results based on the LASSO selection only), and (3) and the naive model refitting approach \cite{belzak2020improving}. The simulation study design and evaluation criteria are discussed in Section \ref{ss:studydesign} and Section \ref{ss:evaluationcriterion}, respectively. All computations are performed in R \cite{R}. The source code will be made publicly available through the Open Science Framework (OSF) upon acceptance of the manuscript.  
\subsection{Study Design}\label{ss:studydesign}
Binary response data are generated from a unidimensional MNLFA model under two conditions\textemdash with or without DIF items. Two factors are manipulated including (1) the total sample size ($n= 500, 1,000, \&\ 2,500$) and (2) the percentage of DIF items ($0\%,\, 25\%,\, \&\ 50\%$). As a result, there are $3 \times 3 = 9$ fully crossed conditions. These manipulating factors are selected due to their relevance to the DIF detection mechanism of the Dscore test. Moreover, actual values of the manipulated factors were chosen to conform to real-world data analytic scenarios and align with previous methodological studies \cite<e.g.,>{bauer2017more,belzak2020improving}. Due to the limited DIF research on the topic, model parameters were generated based on a real data analysis using the UK normative sample data of the Revised Eysenck Personality Questionnaire \cite<EPQ-R, >{eysenck1985revised}\footnote{We are grateful to Dr. Paul Barrett for granting us access to the
data}. True data generating values of these DIF items are described in details in the following section. 
\subsection{The True Data Generation Model}

The binary response for examinee $i$ for item $j$,
denoted $Y_{ij} \in \{0,1\}$, $i= 1,\dots, n$, $j=1,\dots,J$, was generated from a MNLFA model (see Equation \ref{eq:cIRF}). The total number of items was fixed at $J=12$. Model parameters of the data generating model are tabulated in Table \ref{tab:truepar}. Specifically, item discrimination and intercept parameters are the ML estimates of item parameters using the EPQ-R data with five items identified as anchors. Three person covariates are considered in the study which mimic gender, age, and their product. Specially, the dichotomous grouping variable is generated from a Bernoulli distribution with success probability of $.5$ (i.e., $x \sim \hbox{Bern}(0.5)$). Then, for those in category 1, age is generated from $\mathcal{N}(0.2, 1)$. Otherwise, age is generated from $\mathcal{N}(0, 1)$, which creates a correlation between age and gender of 0.1. An interaction effect is then created by multiplying age and gender. By including an interaction effect, the correlations between the interaction effect and the two variables is large ($>.6$). Furthermore, three types of DIF items were generated. Under the $25\%$ DIF condition, items 1 to 3 are DIF items representing items with large, medium, and small DIF effect sizes. When the proportion of DIF items is increased to $50\%$, items 4 to 6 are added to the DIF set and their DIF effect sizes replicate those of items 1-3.
\tablehere{2}

\begin{figure}[htbp!]
 \centering
  \includegraphics[width=0.65\textwidth]{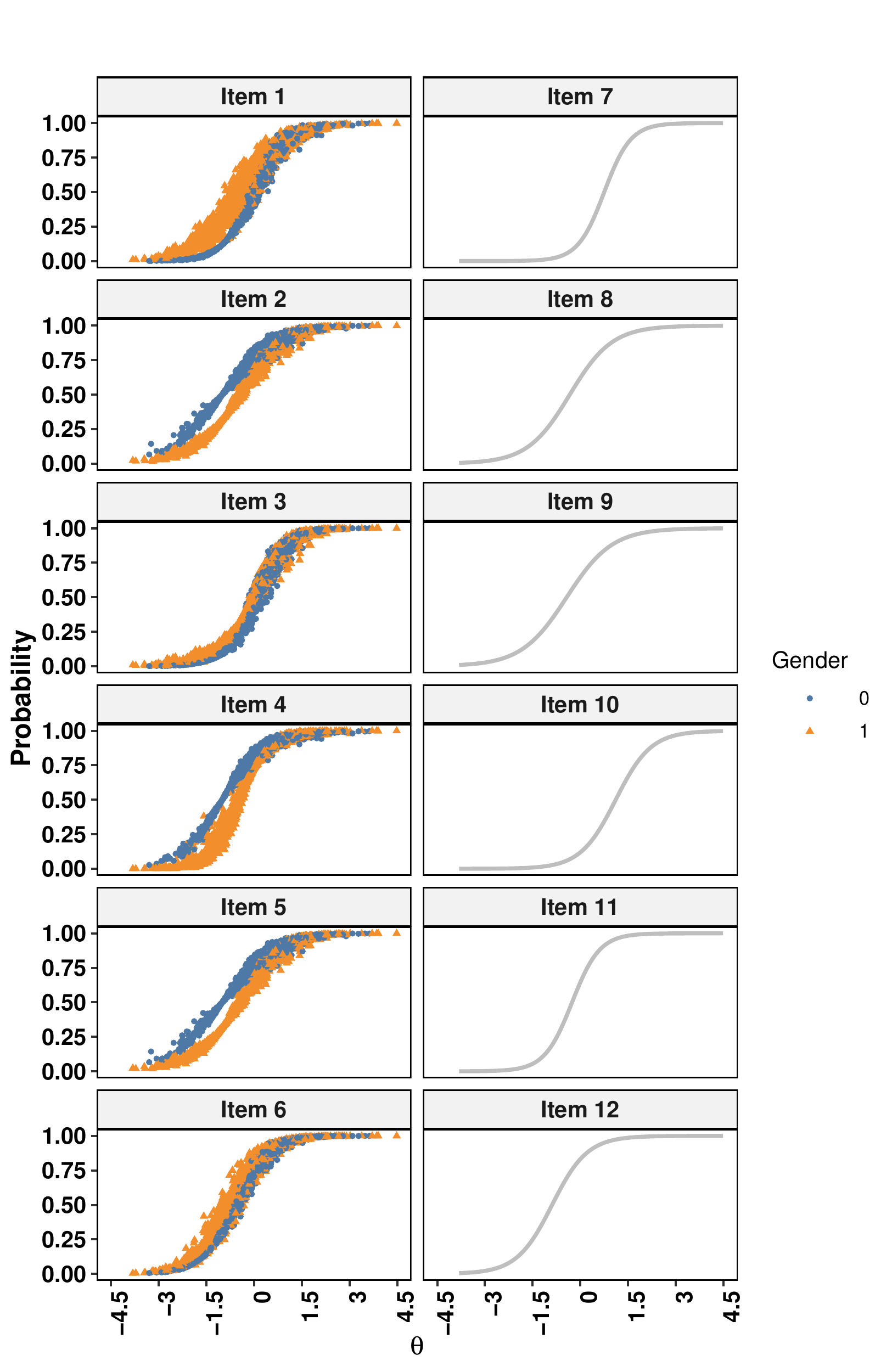}
 \caption[Conditional probability of endorsing an item using the true data generation parameters]{Conditional probability of endorsing an item. $2,500$ item responses to all 12 items were generated using the moderated nonlinear factor analysis model with three covariates. Items 1 to 6 are DIF items.}
  \label{fig:truepar}
\end{figure}

To visualize the generated items, Figure \ref{fig:truepar} displays the true probability of each person endorsing items with the condition of $50\%$ DIF items and $n=2,500$. Plots on the left correspond to DIF items whereas those on the right visualize non-DIF items. Gender influence on the item response can be seen from different colors and shapes. 
\figurehere{1}

Lastly, the latent variable follows a normal distribution ${\cal N}(\ggamma^\top \xx,\ddelta^\top \xx)$ conditional on $\xx$, and similarly true population parameter values are shown in Table \ref{tab:truepar}, which are also enlightened by real data. A total of 500 replications for each condition are implemented ensure that the 95\% normal approximation confidence band at the nominal level $0.05$ to be $ [0.031, 0.069]$.
\subsection{Estimation}
For each replication under each condition, the following four methods are fitted to the binary response data to test DIF at the item level and the parameter level. MNLFA models are estimated using penalized maximum likelihood for the Reg-DIF method and the Dscore method or ML for the refit method and oracle solution. For both estimation methods, integration with respect to the latent variable is approximated by a 49-point Gauss–Hermite quadrature. The four methods are discussed in detail here. 

First, each item is tested for DIF based on LASSO selection only (i.e., the Reg-DIF method). The MNLFA model fitted with all three grouping covariates is estimated using the penalized EM algorithm. To select the optimal tuning parameter under each replication, typically BIC is used in latent variable models. However, the current study fixed $\lambda$ at an estimated optimal value (i.e., $c\sqrt{1/N}$ where the constant $c$ is estimated by $\hat{c} = \sqrt{n}\sum^{R}_{r=1}(\hat{\lambda}_r)/R$)\footnote{A pilot study found that BIC calculated based on the penalized EM parameter estimates always select a relatively smaller $\lambda$ value when the DIF effect size is large, which results in too many FDR. This phenomena persists even when sample size is large $n= 2,500$ or $n=5,000$. We suspect that the less optimal $\lambda$ selection is due to parameter bias due to the regularization. This can be verified that when the $\lambda$ value is selected by the BIC value calculated based on the maximum likelihood estimates (i.e., the model refitting method), the FDR rate is more controlled especially under the large sample size condition.}. The conditions with $n=500$ were used to estimate the constant. Results show that $\hat{c}= 0.8291,0.6883,0.5727$ when the number of DIF items is 0, 3, and 6, respectively. Therefore, $\lambda$ is fixed at the estimated optimal value for each condition. For the reg-DIF method, an item does not exhibit DIF if both $\bbeta_{0j}$ and $\bbeta_{1j}$ are zero vectors. Otherwise, the item is considered as a DIF item.  

Next, the naive model refitting method can be applied by refitting the same MNLFA model with the anchors selected using method 1. This model is estimated using the marginal ML estimation method with the EM algorithm. The marginal likelihood function can be approximated using the same configuration as in the penalized EM algorithm. The convergence tolerance for the log-likelihood change for the penalized EM algorithm and the M-step are set to be $10^{-4}$ and $10^{-6}$, respectively. The maximum number of iterations is set to be $500$. 
Additionally, the Dscore test is conducted after the initial run (method 1) using the penalized EM algorithm. Details of the Dscore test was described in Section \ref{ss:dscoretest} and thus are not repeated here. A critical step to conduct the Dscore test is to estimate $\hat{\WW}$ as displayed in Equation \ref{eq:w}. As the theory only requires that $\lambda$ and $\lambda^\prime$ are of the same rate, $\lambda^\prime$ will be set to the selected $\lambda$ value in the initial parameter estimation step to speed up the computation and save time. \citeA{ning2017general} and \citeA{fang2017testing} have found that the decorrelated score test is not sensitive to $\lambda^\prime$ and both fixed $\lambda^\prime=0.5\sqrt{\log d/n}$ in their simulation studies. Furthermore, asymptotic unbiased parameter estimates can be estimated using the one-step bias correction (see Equation \ref{eq:onestepubiased}). Note that although the focal parameters can be multidimensional or unidimensional (i.e., the debias step can be conducted at the item level or parameter level), the current study investigates the one-step bias correction by parameter type. For instance, for each item, the bias correction treats $(a_j, d_j, \bbeta_{0j}, \bbeta_{1j})^\top$ as the focal parameters and everything else as nuisance parameters. The population parameter estimates are corrected at the same time by treating all item parameters as nuisance parameters. 

Lastly, to compare the performance of the Dscore test with the oracle solution assuming anchors are known, Wald tests are performed to test DIF at the item-level. For each replication and each condition, item 11 and item 12 are treated as anchors. In addition, final model parameter estimates and their corresponding standard errors are estimated using anchors selected by the Wald test. 

\subsection{Evaluation Criteria} \label{ss:evaluationcriterion}
The comparative inferential performance of the four methods is evaluated in terms of (1) hypothesis testing in testing a DIF item measured by rejection rates at the $0.05$ level, (2) parameter recovery including bias and variance of model parameter estimates including a-DIF and d-DIF, and (3) recovery of standard errors. Standard error estimates are evaluated by comparing the square root of the mean of the variance of parameter estimates against the empirical standard errors (i.e., Monte Carlo standard deviation of the parameter estimates).
\section{Results}
\subsection{Results: Hypothesis Testing}
In this section, the Type I error rate, false positive rate (FDR), and power of detecting a DIF item are reported. To visualize the comparative performance, the empirical rejection rate at the nominal level $0.05 $ was plotted under (1) the null condition when there were no DIF items (i.e., Type I error), (2) the alternative condition when there was a mix of DIF and DIF-free items for DIF items only (i.e., Power), and (3) the alternative condition when there was a mix of DIF and DIF-free items for DIF-free items only (i.e., FDR).

\subsubsection{Type I Error Rate}

\begin{figure}[htbp!]
  \centering
   \includegraphics[width=0.7\textwidth]{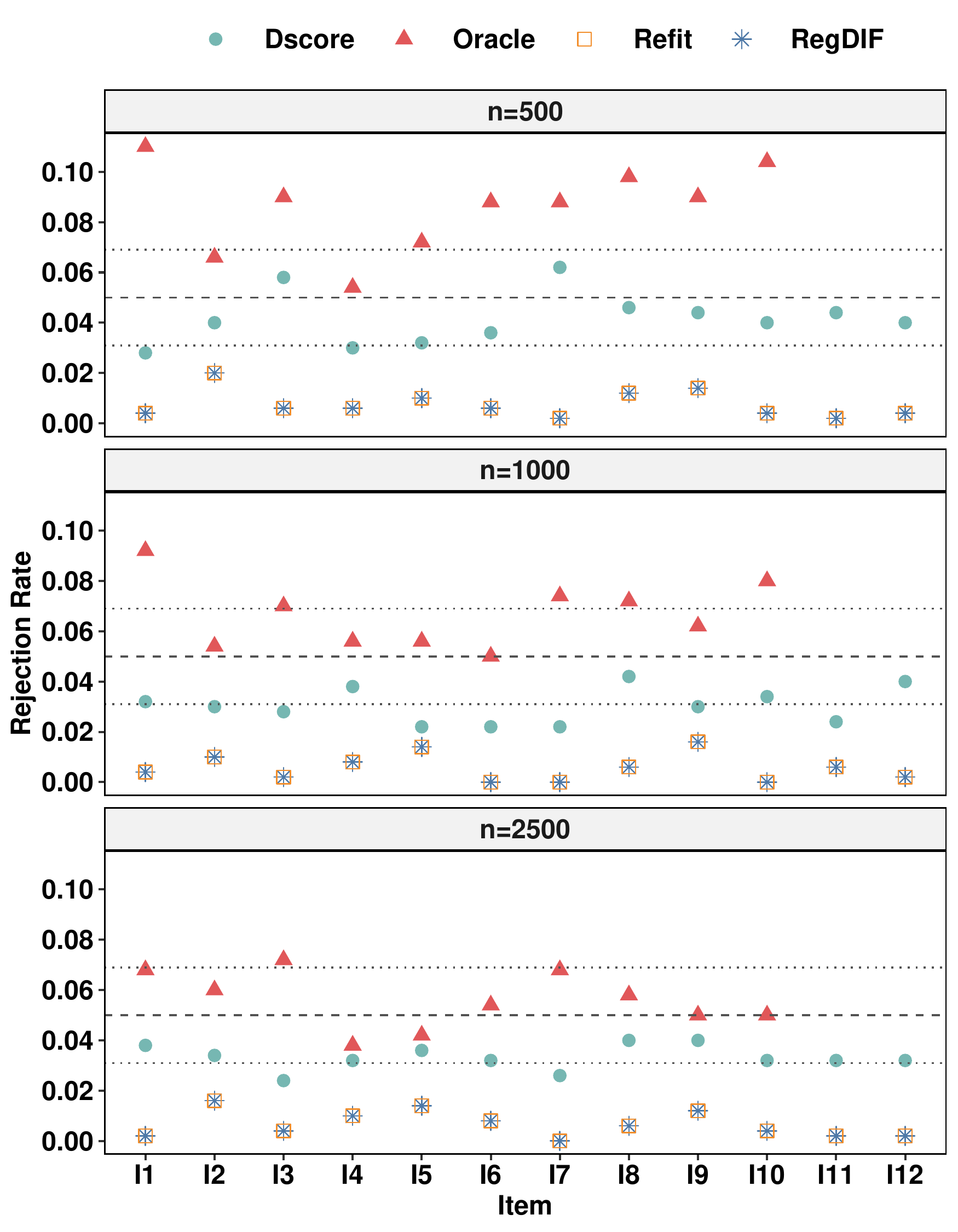}
\caption[Type I error results of incorrectly detecting a DIF item under the null condition]{Type I error results of incorrectly detecting a DIF item under the null condition when there is no DIF items. Different methods are displayed in different colors and shapes. Two dotted horizontal line shows the 95\% normal-approximation confidence band at the nominal level $0.05$ (horizontal dashed line). The oracle solution only performed on item 1 to item 10 as the last two items are treated as anchors.}
  \label{fig:typeierror}
  \vspace{-12pt}
\end{figure}

Figure \ref{fig:typeierror} shows the Type I error rate of detecting a DIF item under the null condition when there were no DIF items. The horizontal dashed lines show the $0.05$ nominal level and the dotted lines represent the 95\% normal approximation confidence band at the nominal level $0.05$ across 500 replications. If the rejection rate falls within the 95\% confidence band (i.e., 95\% confidence interval =$ [0.031, 0.069]$), the specific method is considered to have well-controlled Type I error rate of incorrectly detecting a DIF item. Larger values than the upper bound of the confidence band at the nominal level 0.05 represent over-rejecting the null hypothesis whereas smaller values than the lower bound of the confidence band indicate under-rejecting the null hypothesis. 

Overall, the Reg-DIF, model refit method, and decorrelated score test showed controlled Type I error under all manipulated sample size conditions as the rejection rate of all these methods fall within the 95\% confidence band. 
\figurehere{2}
\subsubsection{Power}

\begin{figure}[htbp!]
  \centering
   \includegraphics[width=1\textwidth]{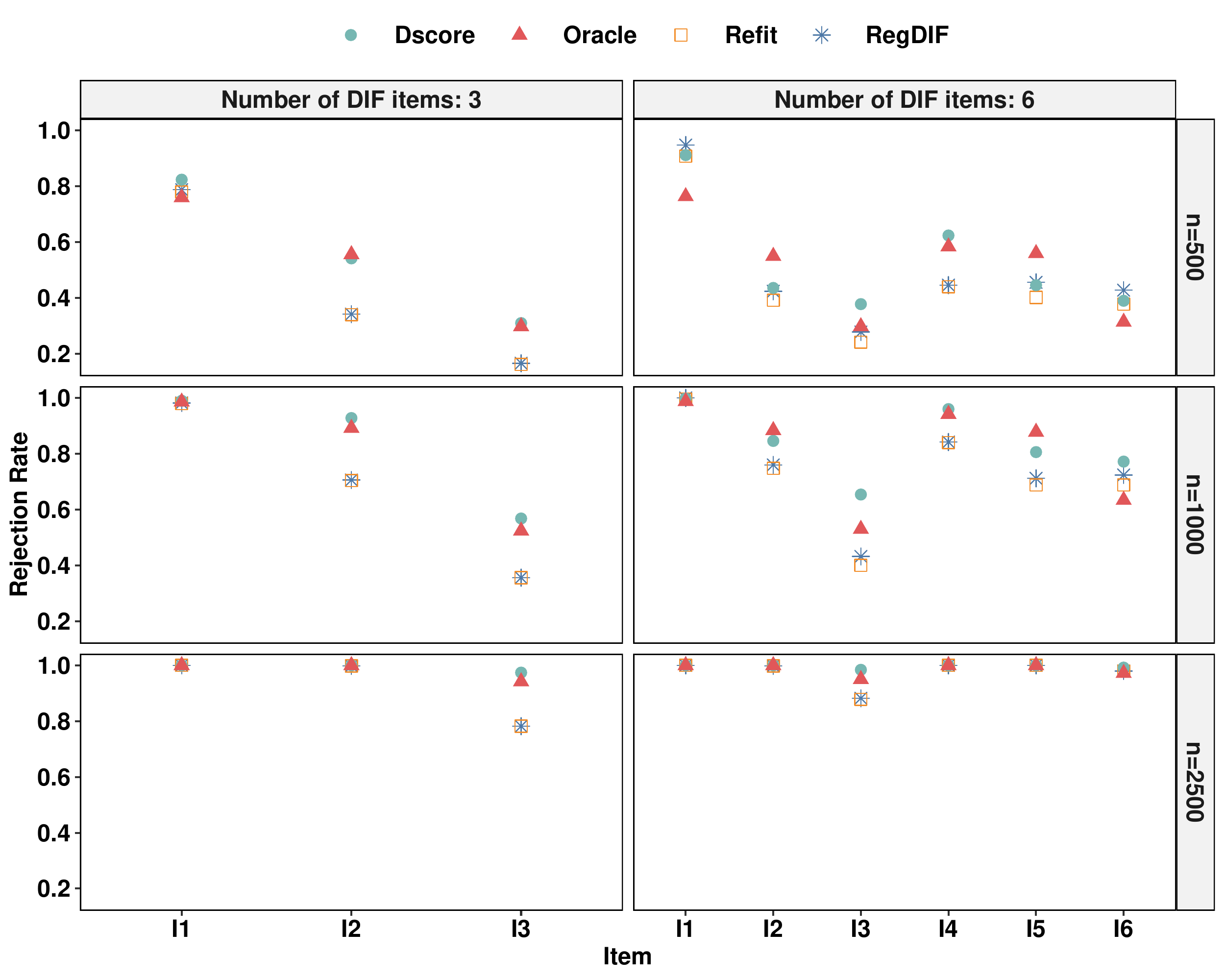}
 \caption[Power results of correctly detecting a DIF item under the alternative condition]{Power results of correctly detecting a DIF item under the alternative condition when there is a mix of DIF and DIF-free items. Different methods are displayed in different colors and shapes. For each method, the empirical rejection rate under the nominal level $0.05$ are calculated for each DIF item under each condition. The column shows the condition when the number of DIF items is 3 or 6 out of 12 items. Each row represents a different sample size condition. The reference dashed line shows the empirical rejection rate = 0.05}
  \label{fig:power}
  \vspace{-12pt}
\end{figure}

Figure \ref{fig:power} shows the power of detecting DIF items under different conditions. The power can be negatively impacted by the sample size while more robust to the number of true DIF items. Overall, the decorrelated score test was more powerful than the Reg-DIF and the model refit methods across all conditions. The difference was more obvious under more challenging conditions where the sample size is small. The performance of the Reg-DIF method and the model refit method was nearly identical when the total number of DIF items was small. The difference between the two methods was more obvious when the number of DIF items was large. Specifically, the model refit method was slightly less powerful than Reg-DIF by design. The reason is that once an item is identified as an anchor item, its DIF parameters are excluded in the refitted model. In other words, this item can no longer be identified as a DIF item.  
\figurehere{3}
\subsubsection{ False Detection Rate}

\begin{figure}[htbp!]
  \centering
   \includegraphics[width=1\textwidth]{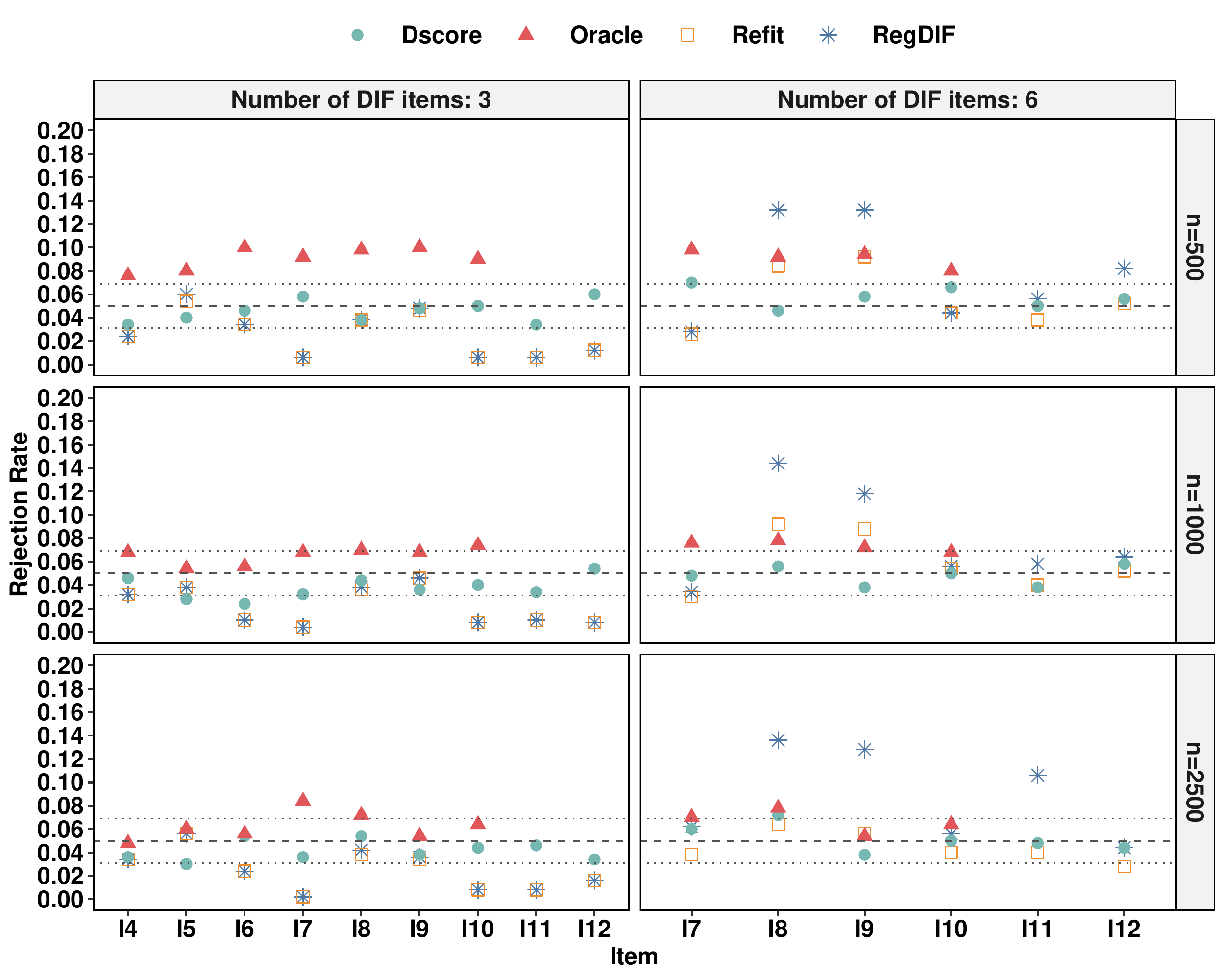}
 \caption[False detection rate results of incorrectly detecting a DIF item under the alternative condition]{False detection rate results of incorrectly detecting a DIF item under the alternative condition when there is a mix of DIF and DIF-free items. Different methods are displayed in different colors and shapes. For each method, the empirical rejection rate at the nominal level $0.05$ are calculated for each DIF item under each condition.  Two dotted horizontal line shows the 95\% normal approximation confidence band for the nominal level $0.05$ (horizontal dashed lines). The column shows the condition when the number of DIF items is 3 or 6 items out of 12 items. Each row represents a sample size condition.}
  \label{fig:fdr}
  \vspace{-12pt}
\end{figure}

Finally, results for incorrectly detecting an anchor item as a DIF item can be visualized using the empirical rejection rate for the true anchor item under alternative conditions. As displayed in Figure \ref{fig:fdr}, the performance of the decorrelated score test is outstanding as the FDR of the decorrelated score test falls within the 95\% normal-approximation confidence band, indicating controlled FDR. Conversely, the Reg-DIF and model refit method under-rejected the null hypothesis when the number of DIF items is small. This is understandable, as was previously shown that the power of identifying the true DIF item was in general smaller than the decorrelated score test and the oracle solution. However, when the number of DIF items was large, the Reg-DIF method over-rejected the null hypothesis. As was previously illustrated, the model refit method is more conservative in rejecting a null hypothesis as compared to the Reg-DIF method. Intuitively, it makes sense that the FDR is more controlled as compared to the Reg-DIF method. Nevertheless, it is still overly rejecting the null hypothesis when the sample size is not sufficiently large. As the sample size increases, the FDR using the model refit method falls within the 95\% confidence band at the nominal 0.05 level.

\figurehere{4}
\subsection{Results: Parameter Recovery}
To investigate parameter recovery, the bias and variance of the parameter estimates using each method were computed. Discussion is focussed on the DIF parameters, the parameters that inferences will be drawn on. Bias and variance of other model parameter estimates including item slope ($a_j$), item intercept ($d_j$), and population parameters ($\ggamma,\ddelta$) using different methods are included in the supplementary materials. For all methods other than the one-step debiased estimator using the decorrelated score function, DIF parameters penalized to be 0 or not included in the final model are treated as 0 to compute bias and variance of an estimator. 
\subsubsection{Bias}

When there was no DIF, all methods recover model parameters well with bias generally less than $0.05$. This is understandable because when there was no DIF, the Type I error rate, as was previously shown, was well-controlled. The penalized EM performed similarly as the EM since in most cases because $\lambda$ was large enough to penalize all DIF parameters to be 0. Even if this does not happen, the model refit method is always fitting the correct model under the null condition.

\begin{figure}[htbp!]
  \centering
   \includegraphics[width=1\textwidth]{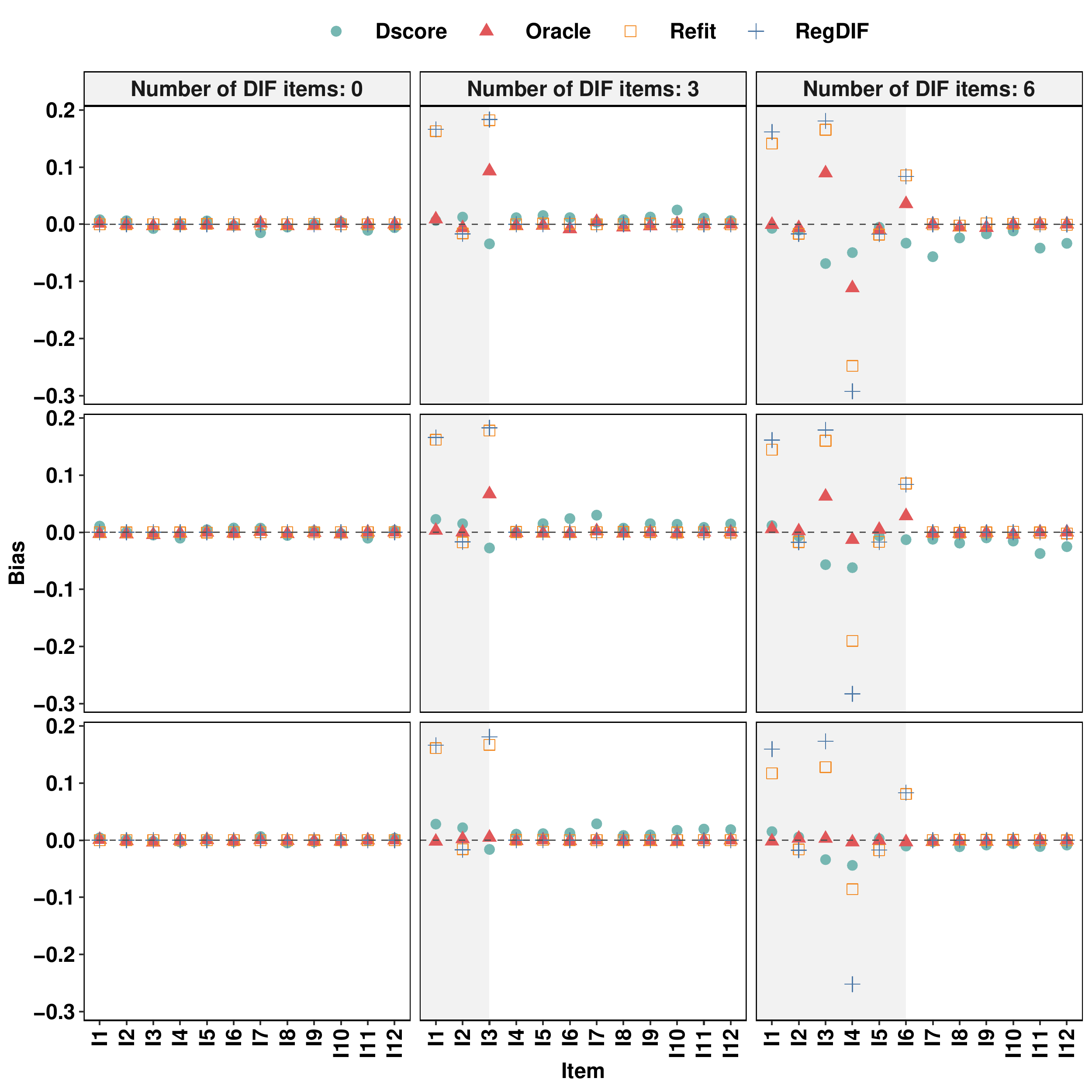}
 \caption[Average bias of a-DIF parameters]{Average bias of a-DIF parameters. Different methods are displayed in different colors and shapes. The column shows the condition when the number of DIF items is 0, 3 or 6 out of 12 items. Each row represents a sample size condition. DIF items are shown in the grey shaded area.}
  \label{fig:bias_ax}
\end{figure}

\begin{figure}[htbp!]
  \centering
   \includegraphics[width=0.7\textwidth]{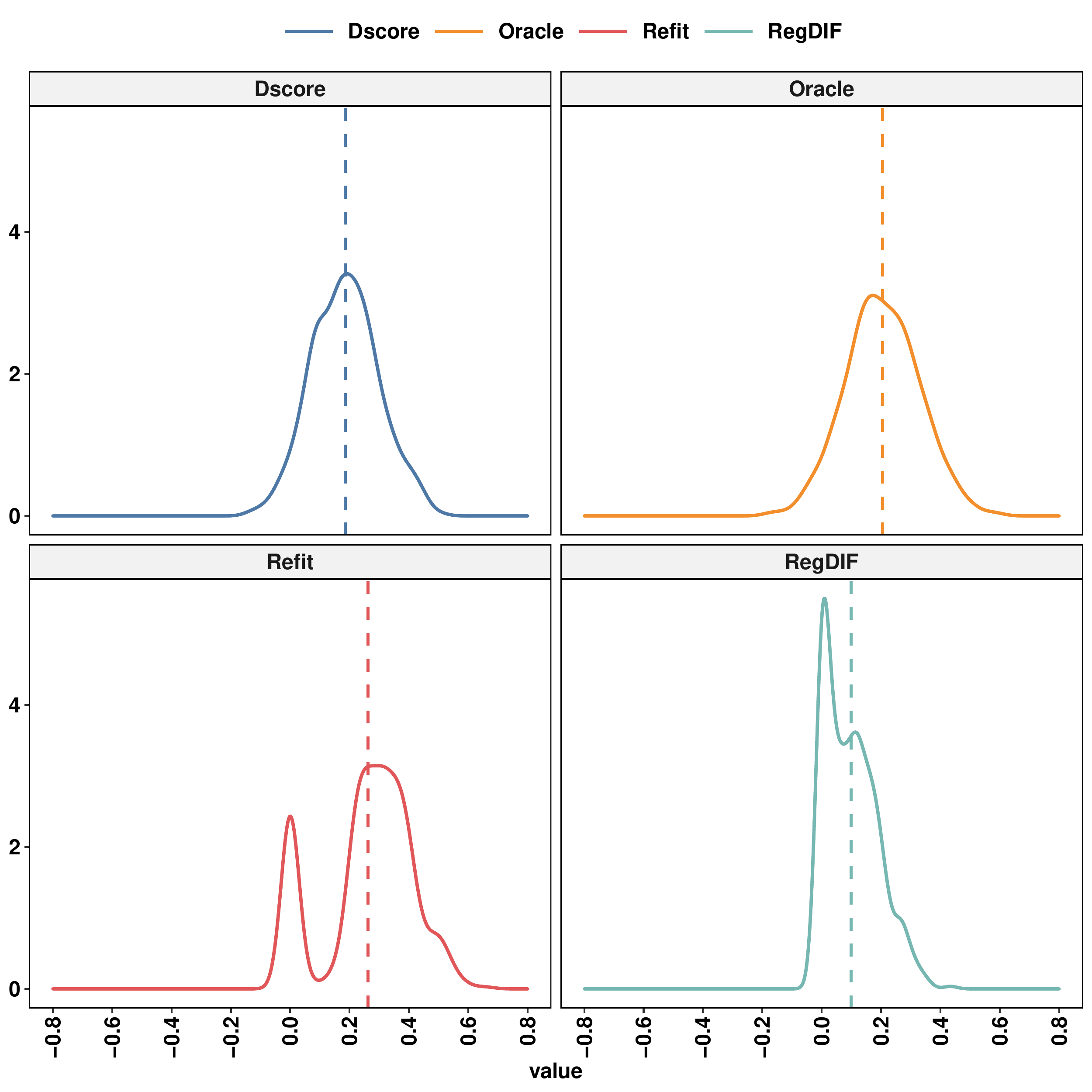}
\caption[Density plot of a non-zero effect]{Density plot of a non-zero DIF parameter by different method. The vertical reference lines indicate the means of the parameter estimates of each effect for each method. The example is plotted using the sample size $2,500$ and first 3-item DIF condition. The non-zero DIF effect is the continuous DIF effect of the first item.}
  \label{fig:samplingdif}
  \vspace{-12pt}
\end{figure}

When there was a mix of DIF and non-DIF items, the performance of different methods varied. The Reg-DIF method often resulted in biased DIF parameters for the true DIF items (i.e., items 1 to 3 in the 3 DIF item condition and item 1 to 6 in the 6 DIF item condition) due to the shrinkage. The bias for the DIF parameters for the anchor items were relatively small. Compared with the Reg-DIF method, the model refit method reduced bias for the true DIF items by $3\%$ to $92\%$ depending on the item. Note that bias remained relatively large for the a-DIF parameter even sample size was large (see last row of the Figure \ref{fig:bias_ax}) due to failing to select the right DIF-effect. It was more clear when the sampling distribution of non-zero DIF parameters was investigated (see Figure \ref{fig:samplingdif}). As can be seen, the sampling distribution of the non-zero DIF effect (blue line) using the model refit method was bi-modal with one mode at 0 indicating that the effect was not selected by the LASSO and another non-zero mode. 

As a comparison, the one-step debiased parameter estimate using the decorrelated score function performed remarkably well. DIF parameters were recovered well as the bias ranged from -0.03 to 0.06 across conditions. Its advantages in reducing bias due to shrinkage were more obvious as compared with the model refit method under smaller sample size or more DIF items conditions. Under the most difficult condition (i.e., small sample size and large number of DIF items), it performed equally well as compared to the oracle solution. More importantly, as the sample size increased, the bias decreased. 
\figurehere{5}
\figurehere{6}
\figurehere{7}

\subsubsection{Variance of DIF Parameter Estimates}

\begin{figure}[htbp!]
  \centering
   \includegraphics[width=1\textwidth]{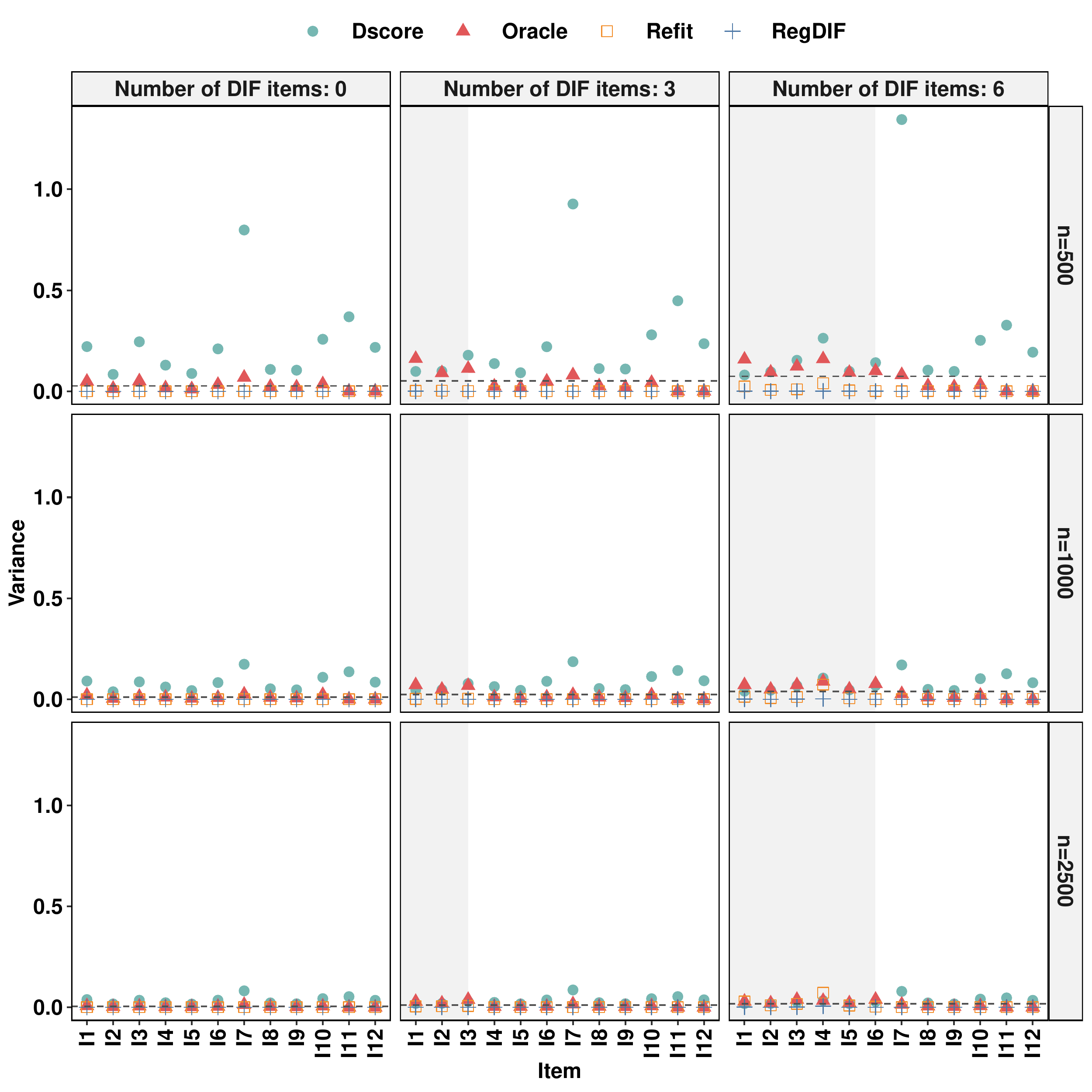}
  \caption[Average variance of a-DIF parameters]{Average variance of a-DIF parameters. Different methods are displayed in different colors and shapes. The column shows conditions when the number of DIF items is 0, 3 or 6 out of 12 items. Each row represents a specific sample size condition. The grey dashed reference line displays the mean variance across all items of the oracle solution for each condition to be used as a benchmark. DIF items are shown in the grey shaded area.}
  \label{fig:Var_ax}
\end{figure}

\begin{figure}[htbp!]
  \centering
   \includegraphics[width=1\textwidth]{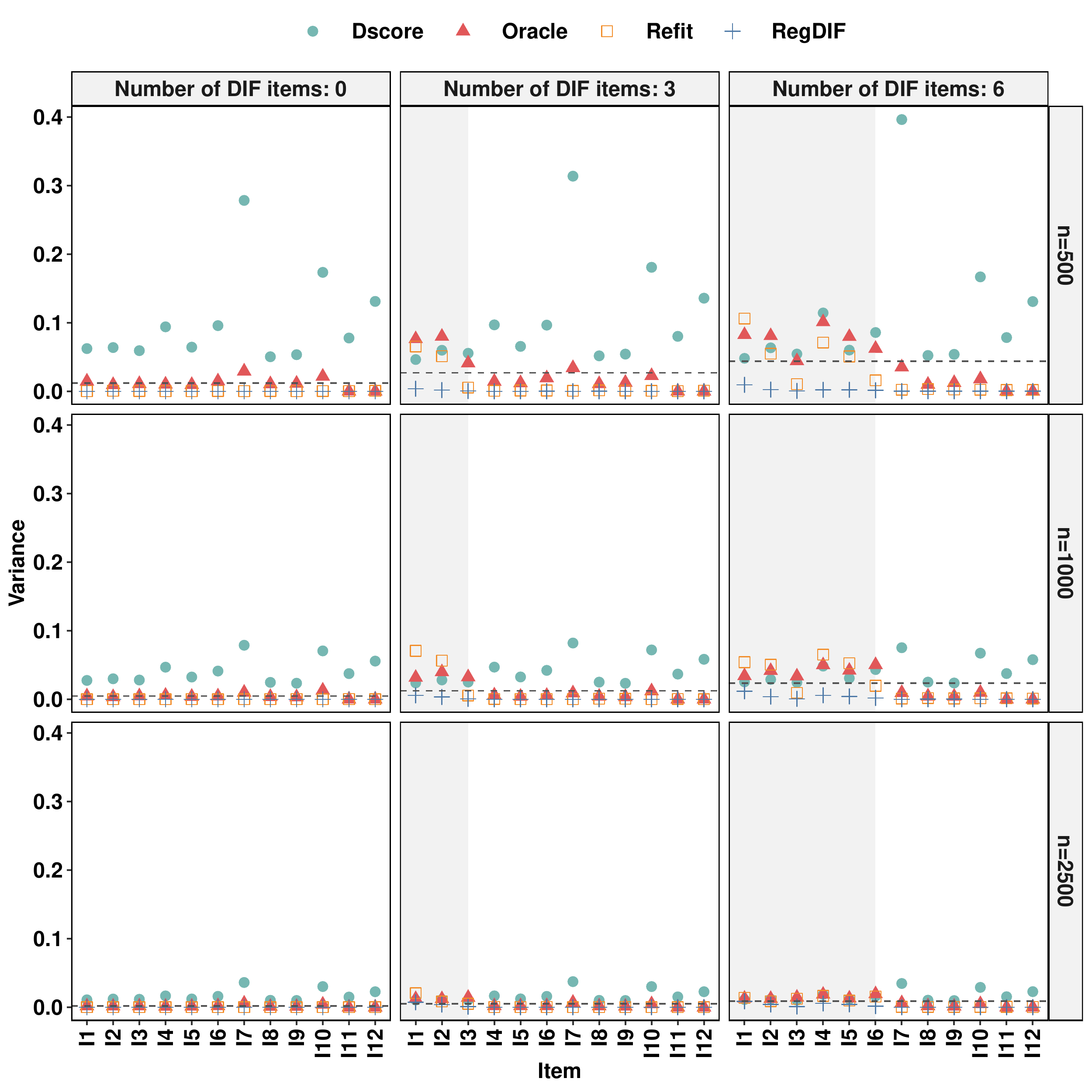}
\caption[Average variance of d-DIF parameters]{Average variance of d-DIF parameters. Different methods are displayed in different colors and shapes. The column shows conditions when the number of DIF items is 0, 3 or 6 out of 12 items. Each row represents a specific sample size condition. The grey dashed reference line displays the mean variance across all items of the oracle solution for each condition to be used as a benchmark. DIF items are shown in the grey shaded area.}
  \label{fig:Var_dx}
\end{figure}

Figure \ref{fig:Var_ax} and Figure \ref{fig:Var_dx} present the variance of the DIF parameter estimates.  Large values indicate more uncertainty of the parameter estimates while smaller values indicate less uncertainty. As the sample size increases, variance should, in general, decrease.
Among all methods, the Reg-DIF method produced the least variable parameter estimates. The model refit method produced similarly variable a-DIF parameter estimates but notably more variable d-DIF parameter estimates as compared to the Reg-DIF especially for the DIF items. Lastly, the one-step debiased parameter estimate using the decorrelated score function was the most variable estimator for the a-DIF parameter as compared to all other methods especially when the sample size was small. However, it was less variable for the d-DIF parameter for the DIF item as compared with the model refit method. 
\figurehere{8}
\figurehere{9}
\subsection{Results: Standard Error Estimates}
The recovery of SE was calculated as the ratio between the square root of the mean of the variance of parameter estimates over the empirical standard deviation of the Monte Carlo parameter estimates. If the ratio is closer to 1, the method has a more valid uncertainty measure. One caveat when calculating the SE using the model refit method is that there is no valid SE estimates for DIF parameters penalized to be 0. If the Reg-DIF method does not select a specific DIF effect for a specific replication, the corresponding DIF covariate is excluded from the model refit method. In such cases, SE estimates of the unselected DIF effect is treated as 0. These arbitrary 0 SEs would artificially decrease the average of the estimated SEs across replications and thus should be interpreted with caution. Alternatively, omitting DIF effects penalized to be 0 will often exclude too many DIF effects across replications resulting in instability of the empirical SEs and inaccurate recovery SE measure\footnote{Recovery of SEs measure for the model refit method based on non-zero DIF effects only has strange large values. Similar findings are reported from \citeA{chen2021advantages}}. 

\begin{figure}[htbp!]
  \centering
   \includegraphics[width=1\textwidth]{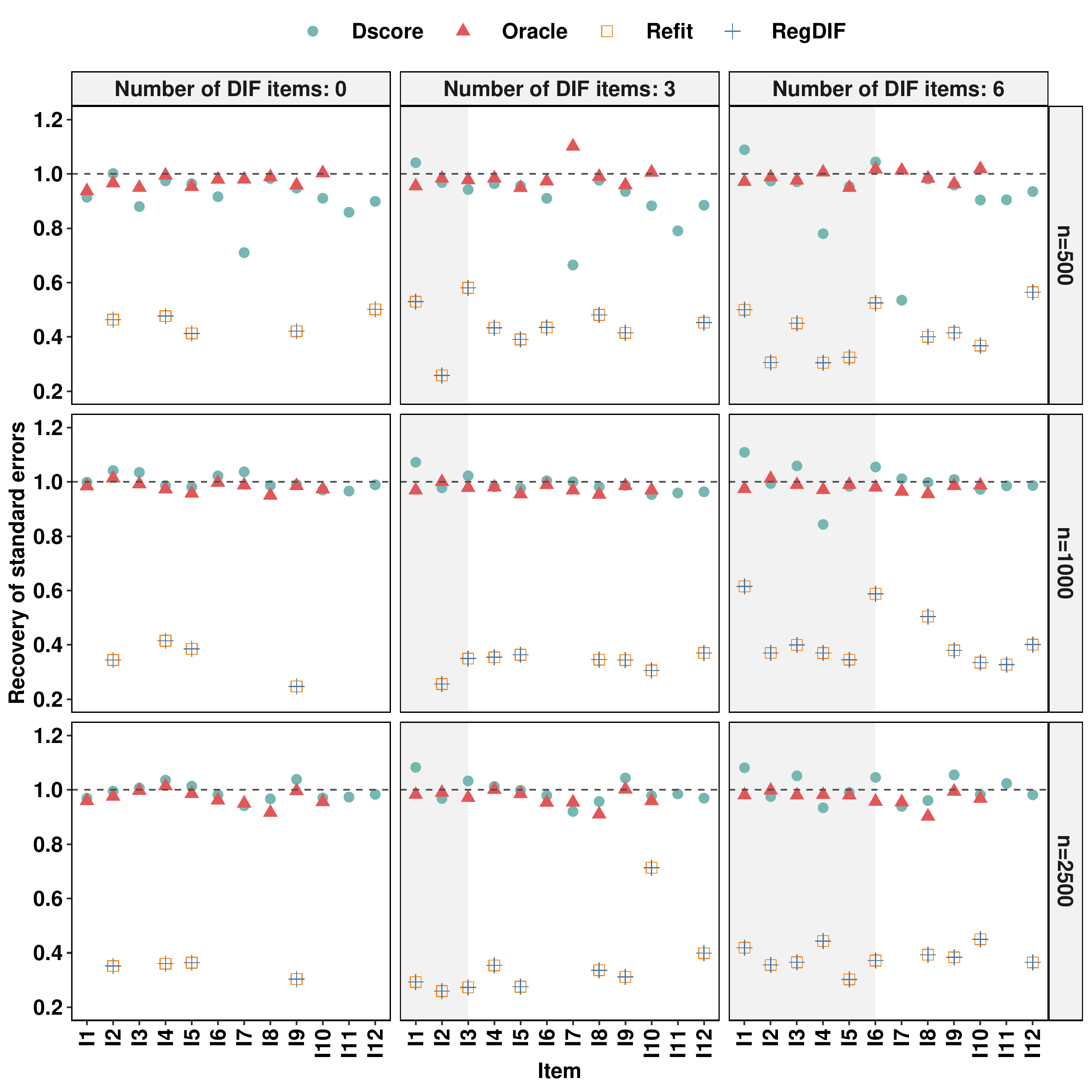}
  \caption[Average standard error recovery of a-DIF parameters]{Average standard error recovery of a-DIF parameters. Different methods are displayed in different colors and shapes. The column shows conditions when the number of DIF items is 0, 3 or 6 out of 12 items. Each row represents a specific sample size condition. The grey dashed reference line displays ratio of 1 indicating perfect recovery.}
  \label{fig:SErecovery_ax}
\end{figure}

\begin{figure}[htbp!]
  \centering
   \includegraphics[width=1\textwidth]{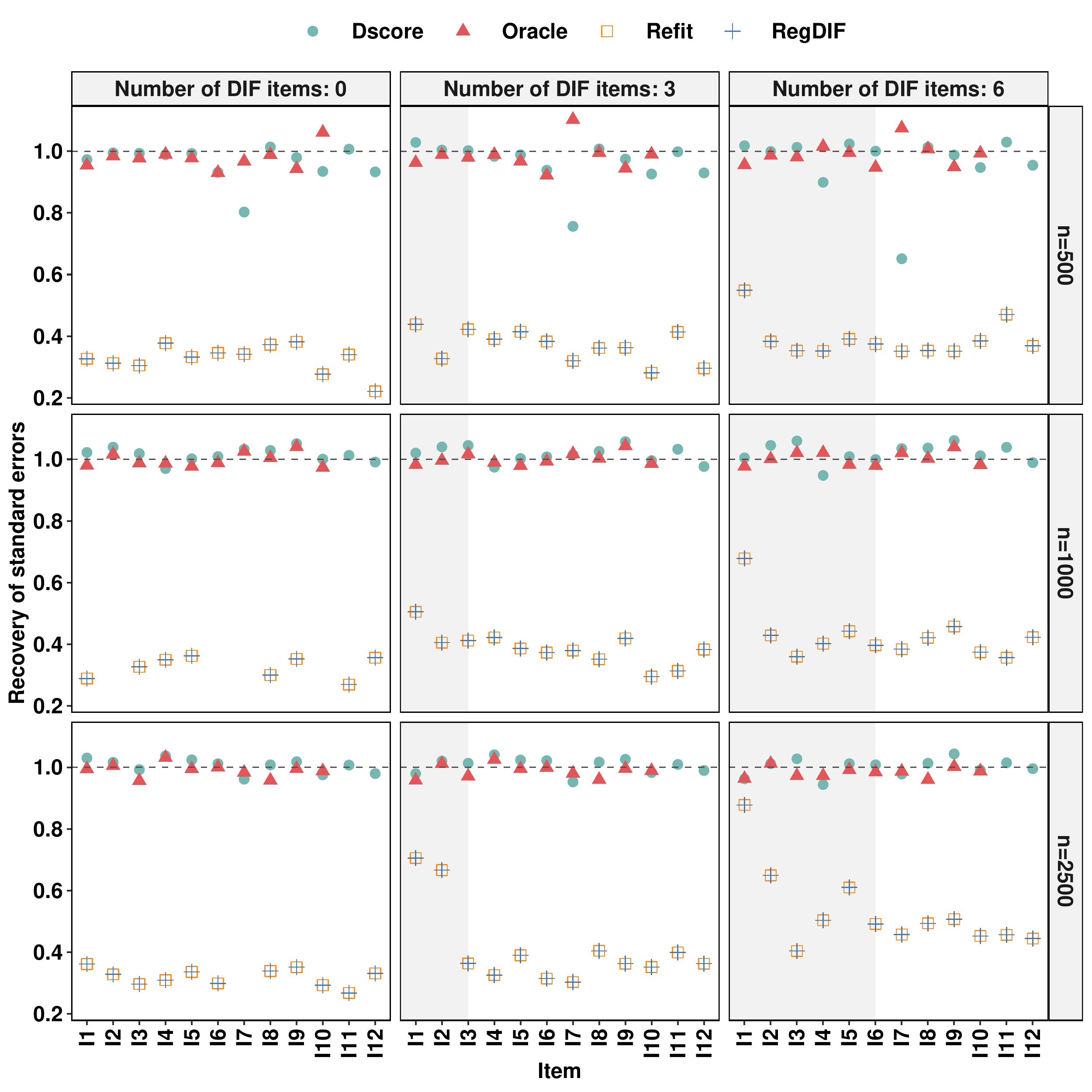}
\caption[Average Standard error recovery of d-DIF parameters]{Average standard error recovery of d-DIF parameters. Different methods are displayed in different colors and shapes. The column shows conditions when the number of DIF items is 0, 3 or 6 out of 12 items. Each row represents a specific sample size condition. The grey dashed reference line displays ratio of 1 indicating perfect recovery.}
  \label{fig:SErecovery_dx}
\end{figure}

The SE recovery for the a-DIF and d-DIF parameters can be visualized in Figure \ref{fig:SErecovery_ax} to \ref{fig:SErecovery_dx}. As there are no valid SE estimates from the penalized EM alogrithm, the Reg-DIF method currently presented in the following figures are based on the SE estimates using the model refit method. Note that as some DIF parameters of certain items were never selected by the Reg-DIF method leaving the recovery of SEs unavailable for the model refit method. This is actually one of the disadvantages of using the model refit method that inference can only be drawn on the selected DIF parameters. Again, the model refit method tended to underestimate SEs of DIF parameters. The underestimation of the SEs of DIF effects could contribute to incorrectly detecting an anchor item as a DIF item as was shown previously shown in Figure \ref{fig:fdr}. As a comparison, the decorrelated score method not only can recover SEs of non-zero DIF effects but also zero DIF effects. For a-DIF and d-DIF, SEs ranged from 0.95 to 1.08 and 0.90 to 1.02, respectively. Note that it appears that SEs of a-DIF and d-DIF effects for item 7 were underestimated. However, given that variances of a-DIF and d-DIF parameters for item 7 were strangely large due to the skewness of the sampling distribution, it makes sense that SEs of DIF parameters of item 7 seem to be underestimated.
\figurehere{10}
\figurehere{11}

\section{Discussion and Conclusion} \label{c:discussion}
\subsection{Summary}
One of the fundamental issues and remaining challenges inherent in DIF detection is finding the correct anchor items. Although recent development in the DIF detection literature (such as using regularization in the MNLFA modeling framework) has shown promising results in detecting DIF items without predefined anchor items, issues such as inflated false detection rate and inability to draw valid inference still remain. 

The goal of the current study is to apply the decorrelated score test to test DIF items based on the $L_1-$ penalized maximum likelihood solution under the MNLFA modeling framework. Unlike DIF detection based on regularization only, the decorrelated score test is valid for all DIF and DIF-free items across all covariates. Additionally, it has been shown that the decorrelated score function can be further used to construct an asymptotically unbiased estimator. Furthermore, the simulation study has shown the comparative performance of the decorrelated score test, DIF detection using regularization only, the model refit method, and the Wald test assuming the correct anchor items in hypothesis testing and parameter recovery. 

Overall, the decorrelated score is advantageous than the other two anchor-free DIF detection methods in three aspects. First, the decorrelated score test shows more statistical power in detecting a DIF item while maintaining controlled false detection rate even compared with the oracle solution with two anchor items available. As summarized and highlighted by previous studies \cite{belzak2020improving,jacobucci2016regularized}, the Reg-DIF method and the model refit method often result in inflated false detection rate in identifying a true DIF item especially when the sample size is small. Similar to these studies, the current simulation study also shows inflated false detection rate especially under the most difficult yet realistic conditions (i.e., when the sample size is small and the number of DIF items is large). As was previously discussed, the current simulation study also shows that the model refit method has similar performance as the decorrelated score test in correctly identifying a true DIF item while controlling for the Type I error when the sample size is large. 

Another important finding from the current simulation study is that the one-step debiased estimator using the decorrelated score function yields less biased item and DIF parameter estimates. Specifically, as compared with the model refit method, the one-step debiased estimator recovers the item slopes, item intercepts, a-DIF, and d-DIF parameters well and sometimes even comparable to the oracle solution. As expected, the model refit method reduces the bias for model parameters estimated from the penalized EM algorithm. However, bias still remains. This is expected as LASSO seems to be more sensitive to detecting d-DIF rather than the a-DIF, a point that we return to at the end of this section, it is likely that a-DIF parameters are likely to be mistakenly unselected. Accordingly, the model refit method fits an incorrect model and, therefore, bias remains. Given the less biased item and DIF parameter estimates, it is recommended to use the one-step debiased estimator to produce reliable point estimates without fitting an additional MNFLA model using EM. However, less bias often indicates more parameter variability. This is evidenced by the fact that there are more uncertainty in the item parameter and DIF parameter estimates using the one-step debiased estimator based on the decorrelated score function. Another interesting observation is that consistent with the model refit method, the one-step debiased estimator always recover item intercept parameters better as compared to item slope parameters (see the supplementary materials). This finding is also aligned other studies \cite<see>[for examples]{bauer2020simplifying, belzak2021using} that d-DIF parameters are in general recovered better than a-DIF parameters, which could potentially explain the superiority of the item intercept parameter recovery. Consequently, a-DIF and item slope parameters have better initial parameter estimates to be plugged into the bias correction function.

Lastly, the decorrelated score test is able to draw reliable statistical inference on a-DIF and d-DIF parameters for zero and non-zero effects, respectively, where the model refit method may fail or is not able to generate one when the DIF effect is zero. As discussed earlier, SEs cannot be computed if DIF effects are penalized to be zero. To this end, the one-step debiased estimator is the only method considered in the present study that can produce reliable statistical inference for the zero DIF effects. As for the remaining DIF effects that are not penalized to be zero, the model refit method often underestimate the SEs of the DIF effects \cite{chen2021advantages,huang2018penalized} leading to incorrect inferences. This may be of concern, if substantive researchers are interested in testing a-DIF and d-DIF separately. If so, the decorrelated score test offers promising and valid statistical test to identify DIF at both the item and the parameter levels.

\section{Future Studies}

Although the proposed decorrelated score test and the one-step debiased estimator have shown promising results in detecting DIF items without predefined anchor items and providing valid uncertainty measures for DIF effects, limitations and issues still remain to be addressed by future studies. 

First, the simulation study has suggested that the one-step debiased estimator does not seem to reduce the bias of the penalized ML estimator and that the associated SEs are unreliably estimated using the efficient information. Given that the ill recovery with the uncertainty measure also happens with unpenalized item parameters, it is conjectured that additional penalty weights might be needed for the population parameter or a more careful selection of penalty weights is needed. Some preliminary investigation suggests that if the population parameter is penalized in the initial $L_1$ penalization stage, the bias correction using the decorrelated score function shows better performance. Future studies are needed to examine the behavior of the decorrelated score function when population parameters are penalized. Alternatively, finding the optimal $\lambda^\prime$ for the population parameter for each iteration might also improve the unpredicted behavior of SEs. In the current simulation study, $\lambda^\prime$ is set to the fixed $\lambda$ based on the estimated rate as the theory suggests $\lambda \asymp \sqrt{1/n}$ and $\lambda^{\prime}\asymp\sqrt{1/n}$ are approximately the same rate to ensure the $l_2$ error bound for initial parameter estimates. It might worth to explore the influence of $\lambda^\prime$ on the SE recovery of the unpenalized parameter. Although \citeA{ning2017general} and \citeA{fang2017testing} have reported that the performance of the decorrelated score test is insensitive to the actual $\lambda^\prime$ value in the high-dimensional regression model and proportional hazard model, it is not clear whether $\lambda^{\prime}$ will impact the one-step debiased estimator especially when there are unpenalized parameters in the model. 

Second, as $\lambda$ is critically important for the initial parameter estimates and the accuracy of DIF detection using the Reg-DIF method and the model refit method, future research is encouraged to investigate different model selection critera in addition to BIC. As mentioned in Section \ref{s:simulation}, a pilot study found that BIC calculated based on the penalized EM parameter estimates always selects a relatively smaller $\lambda$ value when the DIF effect size is relatively large and thus results in too many false positives. The current study used a fixed $\lambda$ value estimated from the rate calculated from the $n=500$ condition based on the ML estimator which potentially avoids the issue of poor performance of BIC. This may also explains the relatively smaller inflated false detection rate and power as compared with other Reg-DIF studies \cite<e.g.,>{bauer2020simplifying,belzak2020improving}. Given the sparse literature on the model selection accuracy in the regularized DIF framework, future research is needed to find viable and efficient penalty selection approaches. 

Third, future research is needed to investigate the impact of different penalty functions on the performance of the decorrelated score test in DIF detection and the asymptotic behavior of the debiased parameter estimator. The general theory provided in \citeA{ning2017general} can be directly applied to many penalty functions \cite<e.g., adaptive LASSO,>[]{zou2006adaptive}. Our simulation study along with the other studies \cite<e.g.,>{belzak2020improving} showed similar differential performance in recovering different parameters and SEs. In addition, applying different penalty weights for a-DIF and d-DIF may yield better initial parameter estimates and thus improve the performance of the one-step debias estimator. 

Lastly, DIF effect size measures can be critically important when deciding to remove, modify, or keep a flagged DIF item. In practice it might be more desirable or more realistic to flag items with large DIF impact, which makes effect size reporting more essential in DIF detection. Latent regression models such as MNLFA models provide a natural DIF effect size measure (i.e., the person covariate effect on the item slope $\bbeta_{1j}$ and item intercept $\bbeta_{0j}$). However, more thorough and careful study on the meaningful cut-off values are needed to facilitate decision making. Alternatively, item level or scale level effects can also be helpful and can be extended to the MNLFA modeling framework. For example, average unsigned difference \cite<see>{woods2011dif} which calculates weighted differences in the expected response functions between the focal and reference groups can be extended to the MNLFA model for selected values or levels of person covariate to evaluate the magnitude of the DIF effect at the item level. In addition to the item level DIF impact, the differential test function (DTF) index \cite{roju1995irt} or the expected total test score difference due to DTF \cite{stark2004examining} can be computed to inform the overall DIF effect on the scale level. 

In sum, the proposed decorrelated score test and its one-step debiased estimator based on the regularized moderated nonlinear factor analysis model offers a promising solution to the practical obstacles encountered by conventional IRT methods. Valid inference at the DIF parameter level opens doors for more complicated substantive research questions such as investigating the complex nature of DIF due to interconnection of background characteristics.

\newpage

\appendix
\renewcommand{\theequation}{A\arabic{equation}}
\setcounter{equation}{0}
\renewcommand{\thesection}{\Alph{subsection}}
\setcounter{section}{0}

\section*{Appendix A}\label{ap:Appendixa}
\subsection{Penalized Expectation-Maximization Algorithm}
The Expectation-Maximization (EM) algorithm \cite{bock1982marginal} is used to find the local minimizer of $ p_n(\xxi; \YY, \lambda|\XX)$ with respect to $\xxi$. The basic idea of the penalized EM algorithm is to repeatedly approximate the upper bound of the penalized negative sample log-likelihood $p_n(\xxi;\YY,\lambda|\XX) \leq q_n(\xxi;\xxi^{(r)}), r=0,1,2,\cdots$ defined in Equation \ref{eq:q} at each iteration (E-step) and obtain the optimizer (M-step). As a result, the alternation between the E-step and the M-step produces a sequence of parameter updates $\xxi^{(r)}$. The final parameter estimates $\hat{\xxi}$ is referred to as the penalized ML estimator. A critical property of the EM algorithm is that the parameter estimate updates the penalized negative sample log-likelihood in a non-increasing fashion (i.e., $p_n(\xxi^{(r+1)};\YY,\lambda|\XX)-p_n(\xxi^{(r)};\YY,\lambda|\XX)\leq q_n(\xxi^{(r+1)},\xxi^{(r)}) - q_n(\xxi^{(r)},\xxi^{(r)})\leq 0$ where $q_n(\xxi,\xxi^{r})$ can be viewed as the upper bound of $p_n(\xxi;\YY,\lambda|\XX)$). Specifically, at iteration $r$, $q_n(\xxi;\xxi^{(r)})$ is defined as
\begin{align} \label{eq:q}
 q_n(\xxi;\xxi^{(r)}) &= \mathbb{E}_{(\theta_i|\xxi^{(r)},\YY,\XX)}\left[p_n(\xxi; \YY;\theta_i|\XX)\right]\\
 &=-\frac{1}{n}\sum_{i=1}^n\mathbb{E}_{(\theta_i|\xxi^{(r)},\yy_i,\xx_i)}\left[\log f(\xxi; \yy_i;\theta_i|\xx_i)\right] +\ssum p_\lambda(\bbeta_j)\\
  &=-\frac{1}{n}\sum_i^{n} \int f(\theta_i|\ggamma^{(r)},\ddelta^{(r)},\yy_i,\xx_i)\biggl(\log \left( f(\yy_i|\theta_i,\xxi,\xx_i)f(\theta_i|\xxi,\xx_i) \right) \biggl) d \theta_i \nonumber\\
  \quad &+\ssum p_\lambda(\bbeta_j)
\end{align}
Dropping $\ssum p_\lambda(\bbeta_j)$ for simplicity, the first term of the right hand side (RHS) equals
\vspace{-12pt}
\begin{align*}
  &\hbox{First term of the RHS}\\
  &=-\frac{1}{n}\sum_{i=1}^{n} \int f(\theta_i|\ggamma^{(r)},\ddelta^{(r)},\yy_i,\xx_i)\biggl(\sum_{j=1}^J\log f_j(\yy_{ij}|\theta_i,\xxi_j,\xx_i)+\log f(\theta_i|\ggamma,\ddelta,\xx_i) \biggl) d \theta_i\\
&=-\frac{1}{n}\sum_{i=1}^{n} \int f(\theta_i|\ggamma^{(r)},\ddelta^{(r)},\yy_i,\xx_i)\biggl(\sum_{j=1}^J\log f_j(\yy_{ij}|\theta_i,\xxi_j,\xx_i) \biggl) d \theta_i \\
&\quad -\frac{1}{n}\sum_{i=1}^{n} \int f(\theta_i|\ggamma^{(r)},\ddelta^{(r)},\yy_i,\xx_i)\biggl ( \log f(\theta_i|\ggamma,\ddelta,\xx_i) \biggl) d \theta_i\\
&=-\frac{1}{n} \sum_{j=1}^J \sum_{i=1}^{n} \int f(\theta_i|\ggamma^{(r)},\ddelta^{(r)},\yy_i,\xx_i)\biggl(\log f_j(\yy_{ij}|\theta_i,\xxi_j,\xx_i) \biggl) d \theta_i\\
&\quad -\frac{1}{n}\sum_{i=1}^{n} \int f(\theta_i|\ggamma^{(r)},\ddelta^{(r)},\yy_i,\xx_i)\biggl ( \log f(\theta_i|\ggamma,\ddelta,\xx_i) \biggl) d \theta_i.
\end{align*}
Further, denote $\theta_{iq}, q=1,\cdots, Q$ and $w_{iq}$ as quadrature nodes and weights for person $i$, respectively. The intractable integral in the $q_n(\xxi;\xxi^{(r)})$ can be approximated by summations on this quadrature grid as 
\begin{align}
	q_n(\xxi;\xxi^{(r)}) &\approx 
	-\frac{1}{n} \sum_{j=1}^J \sum_{i=1}^{n} \sum_{q=1}^Q  e_{iq}^{(r)}\biggl(\log f_j(\yy_{ij}|\theta_{iq},\xxi_j,\xx_i) \biggl) +\sum_{j}^J p_\lambda(\bbeta_j) \\
&\quad -\frac{1}{n} \sum_{i=1}^{n} \sum_{q=1}^Q  e_{iq}^{(r)}\biggl ( \log f(\theta_{iq}|\ggamma,\ddelta,\xx_i) \biggl),
\end{align}
where $ e_{iq}^{(r)}=f(\theta_i|\ggamma^{(r)},\ddelta^{(r)},\yy_i, \xx_i)$ is the posterior probability of $\theta_i$ at iteration $r$, which is approximated by 
\begin{align} \label{eq:ewt}
  e_{iq}^{(r)} = \frac{\prod_{j=1}^J f_j(y_{ij}|\theta_{iq},\xxi_j^{(r)},\xx_i)w_{iq}}{\sum_{q'=1}^	Q\prod_{j=1}^J f_j(y_{ij}|\theta_{iq'},\xxi_j^{(r)},\xx_i)w_{iq'}}.
\end{align}
Finally, the Bock-Aitkin EM algorithm with the $L_1$ penalty can be achieve using the following steps until
convergence is reached:
\begin{itemize}
  \item[] \textsc{E-step}. Compute the posterior weights $e_{iq}^{(r)}$, $i =
    1,\dots, n$, $q = 1,\dots, Q$;
  \item[] \textsc{M-step}.
  \begin{itemize}
   \item[] For latent population parameters, compute 
   \begin{align}\label{eq:msteplatent}
   (\ggamma^{(r + 1)},\ddelta^{(r+1)})^\prime =
    \arg\min_{\ggamma,\ddelta}\Biggl\{-\frac{1}{n}
    \sum_{i=1}^{n} \sum_{q=1}^Q  e_{iq}^{(r)}\biggl ( \log f(\theta_{iq}|\ggamma,\ddelta,\xx_i) \biggl)\Biggr\}
   \end{align}

   \item[] For parameters for item $j$, compute 
   \begin{align} \label{eq:mstepitem}
   \xxi_j^{(r + 1)} =
    \arg\min_{\xxi_j}\Biggl\{-\frac{1}{n}
    \sum_{i=1}^{n} \sum_{q=1}^Q  e_{iq}^{(r)}\biggl(\log f_j(y_{ij}|\theta_{iq},\xxi_j,\xx_i) \biggl) + p_\lambda(\bbeta_j)\Biggr\}.
   \end{align}   
     \end{itemize}
\end{itemize}
\subsection{M-step Optimization}
As was previously shown, at each iteration $r+1$, the M-step optimizes $\xxi^{(r+1)}=\arg\min_{\xxi} p_n (\xxi;\YY,\lambda|\XX)$, which can be split into two optimization problems shown in Equations \ref{eq:msteplatent} and \ref{eq:mstepitem}. These two equations update the population parameters (i.e., $\ggamma$ and $\ddelta$) and item parameters, respectively. Equation \ref{eq:msteplatent} can be obtained by a Newton-type optimizer. The Broyden–Fletcher–Goldfarb–Shanno (BFGS) algorithm is used in the current study. In contrast, Equation \ref{eq:mstepitem} for each item $j$ needs to be handled separately due to the non-differentiability of the $L_1$ penalty. Finding $\xxi_j^{(r+1)}$, $j=1, \dots, J$ in the M-step amounts to a conditional density estimation problem in a sample size $nQ$ with weights $e_{iq}^{(r)}$. Specifically, maximizing the unpenalized conditional density can be achieved by solving the reweighted least square problem. With the $L_1$ penalty, the coordinate descent algorithm \cite{friedman2010} can be used to solve the penalized weighted least square problem with pseudo data of a sample size $nQ$.  

\newpage
\section*{Appendix B} \label{ch:Appendixb}

As mentioned before, the general theory of the decorrelated score test and its one-step estimator are established under some assumptions \cite{ning2017general}. These assumptions need to be verified to guarantee the asymptotic properties of the Dscore test and the asymptotic unbiased estimator as shown in Equations \ref{eq:Fdscore} and \ref{eq:Funbiased}. The following four assumptions are briefly summarized and the verification will be provided in the final dissertation.
\begin{assumption}[Consistency conditions for initial parameter estimates]\label{asp:1}
For some sequence $a_1(n)$ and $a_2(n)$ converge to $0$ as $n\rightarrow \infty$, it holds 
\begin{align}
\lim_{n\to\infty} P_{\xxi^*}(\|\hat{\xxi}-\xxi^*\|_1 &\lesssim a_1(n))	=1\\
	\lim_{n\to\infty} P_{{\xxi}^*}(\|\hat{{\WW}}-{\WW}^*\|_1  &\lesssim a_2(n))	=1,
\end{align}
\end{assumption}
\noindent where $\| \|_1$ stands for the $L_1$ operator norm of a matrix (e.g., $\| \mathbf{A}\|_1=\max_{1\leq j\leq m} \sum_{i=1}^n|a_{ij}|$) and $\lesssim $ denotes that the left side is less than or equal to the right hand side times some constant $C>0$.
Although estimation consistency of LASSO has been studied in linear and generalized linear models \cite{knight2000asymptotics,ning2017general}, it has never been studied in the latent variable modeling framework. As the loss function or the negative sample log-likelihood of MNLFA model is not strictly convex, the proof of parameter estimation consistency can be non-trivial. Moreover, as mentioned by \citeA{ning2017general}, there can be extra difficulty in bounding $\|\hat{{\WW}}-{\WW}^*\|$, as $\hat{{\WW}}$ depends on $\hat{\xxi}$.

\begin{assumption}[Concentration of the gradient and Hessian]\label{asp:2}
Let ${\bf V}^*=({\bf I}_{d_0\times d_0}, -{\WW^{*\top}})^\top$, then assume 
\begin{align}
	\|\nabla\ell(\xxi^*)\|_{\infty}&=\bigO_p(\sqrt{\log d/n})\\
	\|\vv^{*\top}\nabla^2\ell(\xxi^*)-\E_{\xxi^*}(\vv^{*\top}\nabla^2\ell(\xxi^*))\|_\infty &=\bigO_p(\sqrt{\log d/n}),
\end{align}
where $\|\|_{\infty}$ of a vector $\vec{A}$ for example indicates $\|\vec{A}\|_\infty = \max_{1\leq i\leq d}|a_i|$ and of a matrix is the maximum absolute row sum of the matrix (i.e., $\|{\mathbf A}\|_\infty = \max_{1\leq i \leq n}\sum_{j=1}^m|a_{ij}|$ ). In the low dimensional setting, it might be sufficient to prove under some finite moment assumptions on the gradient and Hessian matrix. 
\end{assumption}

\begin{assumption}[Local smoothness on the loss function]\label{asp:3}
Let $\hat{\xxi}_0=({\bf 0},\hat{\eeta}^\top)^\top$, $\hat{\vv}=({\bf I}_{d_0\times d_0} ,-\hat{\WW}^\top)^\top$, and ${\vv}^*=({\bf I}_{d_0\times d_0},-{\WW}^{*\top})^\top$. For both $\check{\xxi}=\hat{\xxi}_0$ and  $\check{\xxi}=\hat{\xxi}$, it holds that 
\begin{align}
	\|\vv^{*\top}\{\nabla\ell(\check{\xxi})-\nabla\ell({\xxi}^*)-\nabla^2\ell({\xxi}^*)(\check{\xxi}-\xxi^*)\}\|_{\infty}&=o_p(n^{-1/2})\\
	\|(\hat{\vv}-\vv^*)^\top(\nabla\ell(\check{\xxi})-\nabla\ell({\xxi}^*))\|_{\infty}&=o_p(n^{-1/2}).
\end{align}
\end{assumption}
\noindent This assumption implicitly assumes that the $\ell(\xxi)$ is second-order differentiable. The verification of the assumption should be straightforward if the loss function is a quadratic form of $\xxi$.

\begin{assumption}[Convergence of the score function]\label{asp:4}
Let $\SSigma^*=\lim_{n\to\infty} \hbox{Var}(n^{1/2}\nabla\ell(\xxi^*))$. Then the score function holds that 
\begin{align}
\sqrt{n}\nabla\ell(\xxi^*)^\top\vv^{*}{(\vv^{*\top}\SSigma^*\vv^*)}^{-1}\vv^{*\top}\nabla\ell(\xxi^*)\xrightarrow{\mathcal{D}}{\chi}^2_{d_0}.
\end{align}
\end{assumption}
\noindent This assumption can be established by verifying the Lindeberg's condition, which is a sufficient condition for a sequence of independent random variables. 



\bibliographystyle{apacite}

\vspace{\fill}\pagebreak
\linespacing{1}
\section*{Figures}


\begin{figure}[htbp!]
  \centering
   \includegraphics[width=1\textwidth]{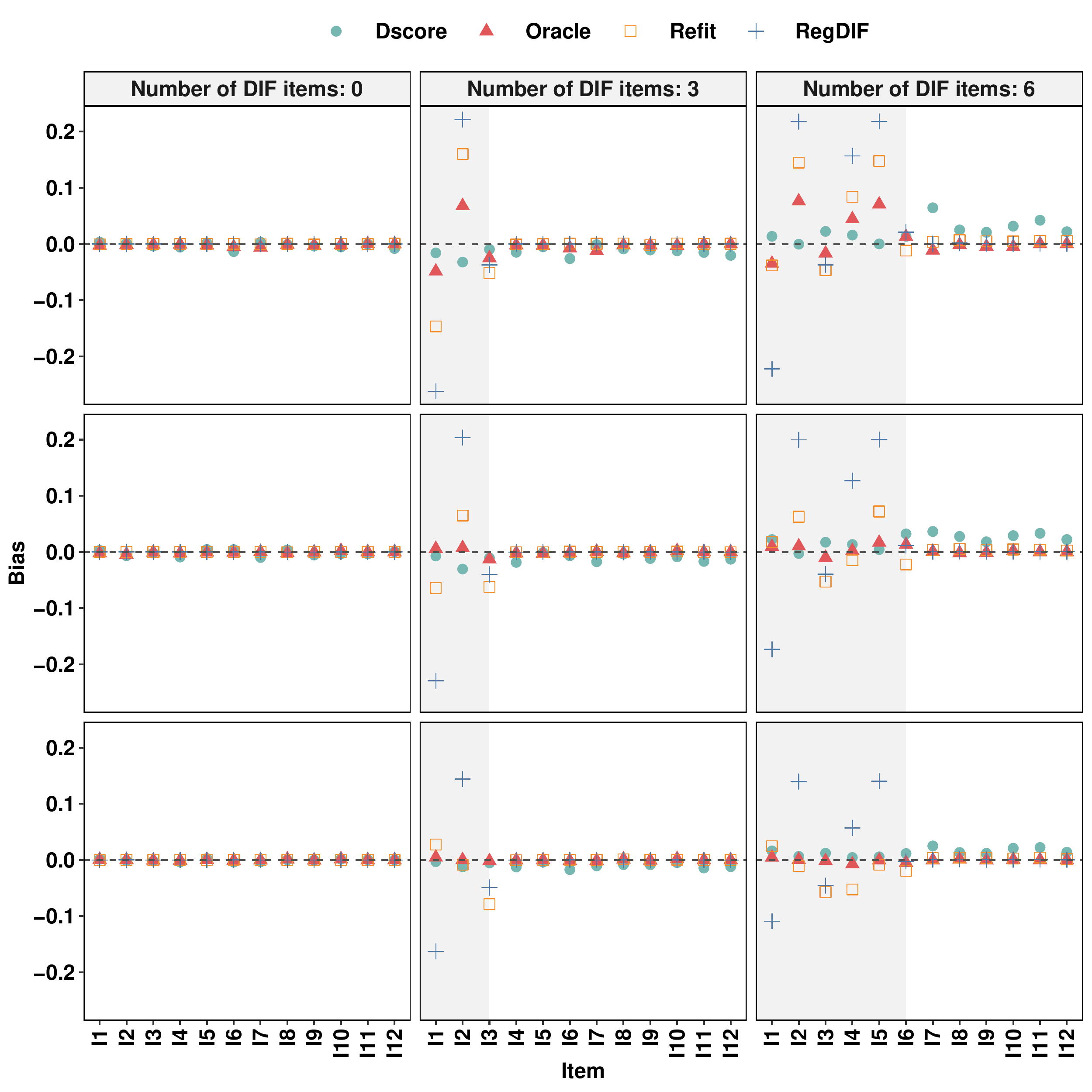}
\caption[Average bias of d-DIF parameters]{Average bias of d-DIF parameters. Different methods are displayed in different colors and shapes. The column shows the condition when the number of DIF items is 0, 3 or 6 out of 12 items. Each row represents a sample size condition. DIF items are shown in the grey shaded area.}
  \label{fig:bias_dx}
\end{figure}

\vspace{\fill}\pagebreak

\section*{Tables}
\begin{table}[ht!]
\caption{Algorithm 1: Decorrelated Score Function Estimation} \label{tab:algorithm1}
\centering
\begin{algorithm}[H]
\caption{Estimated the decorrelated score function}\label{alg:dscore}
\begin{algorithmic}[1]
\Require Negative sample log-likelihood $\ell(\ppsi,\eeta)$, penalty function $p_\lambda()$, and tuning parameters $\lambda$ and $\lambda^\prime$.

\State Estimate $\hat{\xxi}$ using penalized ML as in Equation \ref{eq:pmlikn} and partition $\hat{\xxi}$ into $\hat{\xxi}=(\hat{\ppsi}^\top,\hat{\eeta}^\top)^\top$
\State Estimate $\WW$ column by column 
 \begin{align} \label{eq:w}
 	\hat{\WW}_{*j}=\argmin_{{\bf w}_j}\frac{1}{2n}\sum_i^n\{\nabla_{\ppsi_j}\ell_i(\hat{\xxi})-\WW_{*j}^\top\nabla_{\eeta}\ell_i(\hat{\xxi})\}^2+p_{\lambda^\prime}(\WW_{*j})
 \end{align}
\State Calculate the estimated descorrelated score function using 
 \begin{align}
 	\hat{\ss}(\ppsi,\hat{\eeta}) & = \nabla_{\ppsi}\ell(\ppsi,\hat{\eeta})-\hat{\WW}^\top\nabla_{\eeta}\ell(\ppsi,\hat{\eeta})
 \end{align}
Return $\hat{\ss}(\ppsi,\hat{\eeta})$
\end{algorithmic}
\end{algorithm}
\end{table}

\begin{table}[ht!]
\caption{Model Parameters of the True Data Generating Model} \label{tab:truepar}
\centering
\begin{tabular}{C{1cm} C{1cm} C{1cm} C{1.5cm}C{1.5cm}C{1.5cm}C{1.5cm}C{1.5cm}C{1.5cm}} 
\toprule
 & & &\multicolumn{3}{c}{a-DIF} &  \multicolumn{3}{c}{d-DIF} \\
 & & & Age & Gender & Product  &  Age & Gender & Product \\
\raisebox{2ex}{Item} & 
\raisebox{2ex}{$d_j$}&
\raisebox{2ex}{$a_j$}& $\beta_{11}$& $\beta_{12}$& $\beta_{13}$& $\beta_{01}$& $\beta_{02}$&$\beta_{03}$\\
\hline
1 & 0.00 & 2.00 & 0.20 & 0.50 & 0.20 & 0.20 & -0.50 & -0.20 \\ 
2 & 1.20 & 1.20 & -0.20 & -0.50 & 0.00 & -0.20 & 0.25 & 0.00 \\ 
3 & -0.20 & 2.00 & -0.25 & 0.25 & 0.10 & -0.15 & -0.25 & -0.15 \\ 
$4^*$ & 1.50 & 1.50 & 0.20 & -0.50 & -0.20 & 0.20 & 0.50 & 0.20 \\ 
$5^*$ & 1.20 & 1.20 & -0.20 & -0.50 & 0.00 & -0.20 & 0.25 & 0.00 \\ 
$6^*$ & 1.10 & 1.90 & -0.25 & 0.25 & -0.10 & -0.15 & -0.25 & 0.15 \\ 
7 & -1.80 & 2.40 & 0.00 & 0.00 & 0.00 & 0.00 & 0.00 & 0.00 \\ 
8 & 0.50 & 1.50 & 0.00 & 0.00 & 0.00 & 0.00 & 0.00 & 0.00 \\ 
9 & 0.60 & 1.40 & 0.00 & 0.00 & 0.00 & 0.00 & 0.00 & 0.00 \\ 
10 & -2.00 & 1.80 & 0.00 & 0.00 & 0.00 & 0.00 & 0.00 & 0.00 \\ 
11 & 0.60 & 2.30 & 0.00 & 0.00 & 0.00 & 0.00 & 0.00 & 0.00 \\ 
12 & 1.60 & 1.80 & 0.00 & 0.00 & 0.00 & 0.00 & 0.00 & 0.00 \\ 
$\ggamma$& \multicolumn{8}{c}{$(-0.2,-0.2,-0.2)^\top$} \\
$\ddelta$& \multicolumn{8}{c}{$(-0.1, 0.3, 0.1)^\top$} \\
\bottomrule
\end{tabular}\\
\vspace*{6pt}
{\raggedright \textit{Note. * indicates that effect sizes of these items under the 25\% DIF condition are 0.} }
\end{table}

\end{document}